\documentclass[journal=jacsat,manuscript=article]{achemso}

\usepackage[version=4]{mhchem} 
\usepackage[table]{xcolor} 
\usepackage{float}
\setkeys{acs}{maxauthors=20, etalmode=truncate}

\usepackage[T1]{fontenc} 

\author{Nicholas P. Sloane}
\affiliation[UNSW]
{ARC Centre of Excellence in Exciton Science, School of Physics, University of New South Wales, Sydney, NSW 2052, Australia}
\author{Damon M. de Clercq}
\affiliation[UNSW]
{ARC Centre of Excellence in Exciton Science, School of Chemistry, University of New South Wales, Sydney, NSW 2052, Australia}
\author{Md Arafat Mahmud}
\affiliation[USYD]{School of Physics, The University of Sydney, Sydney, NSW 2006, Australia}
\alsoaffiliation[NANO]{Sydney Nano, The University of Sydney, Sydney, NSW 2006, Australia}

\author{Jianghui Zheng}
\affiliation[USYD]
{School of Physics, The University of Sydney, Sydney, NSW 2006, Australia}
\alsoaffiliation[NANO]{Sydney Nano, The University of Sydney, Sydney, NSW 2006, Australia}
\altaffiliation{Australian Centre for Advanced Photovoltaics (ACAP), School of Photovoltaic and Renewable Energy Engineering, University of New South Wales, Sydney, NSW 2052, Australia}

\author{Adrian Mena}
\affiliation[UNSW]
{ARC Centre of Excellence in Exciton Science, School of Physics, University of New South Wales, Sydney, NSW 2052, Australia}

\author{Michael P. Nielsen}
\affiliation[UNSW]
{School of Photovoltaic and Renewable Energy Engineering, University of New South Wales, Sydney, NSW 2052, Australia}

\author{Anita W.Y. Ho-Baillie}
\affiliation[USYD]
{School of Physics, The University of Sydney, Sydney, NSW 2006, Australia}
\alsoaffiliation[NANO]{Sydney Nano, The University of Sydney, Sydney, NSW 2006, Australia}
\altaffiliation{Australian Centre for Advanced Photovoltaics (ACAP), School of Photovoltaic and Renewable Energy Engineering, University of New South Wales, Sydney, NSW 2052, Australia}

\author{Christopher G. Bailey}
\affiliation[USYD]
{School of Physics, The University of Sydney, Sydney, NSW 2006, Australia}
\alsoaffiliation[NANO]{Sydney Nano, The University of Sydney, Sydney, NSW 2006, Australia}

\author{Timothy W. Schmidt}
\affiliation[UNSW]
{ARC Centre of Excellence in Exciton Science, School of Chemistry, University of New South Wales, Sydney, NSW 2052, Australia}
\author{Dane R. McCamey}
\affiliation[UNSW]
{ARC Centre of Excellence in Exciton Science, School of Physics, University of New South Wales, Sydney, NSW 2052, Australia}
\email{dane.mccamey@unsw.edu.au}

\title[An \textsf{achemso} demo]
  {Mitigating Singlet Exciton Back-Transfer using 2D Spacer Layers for Perovskite-Sensitised Upconversion}

\begin{document}

\begin{tocentry}





\includegraphics[width=8.3cm]{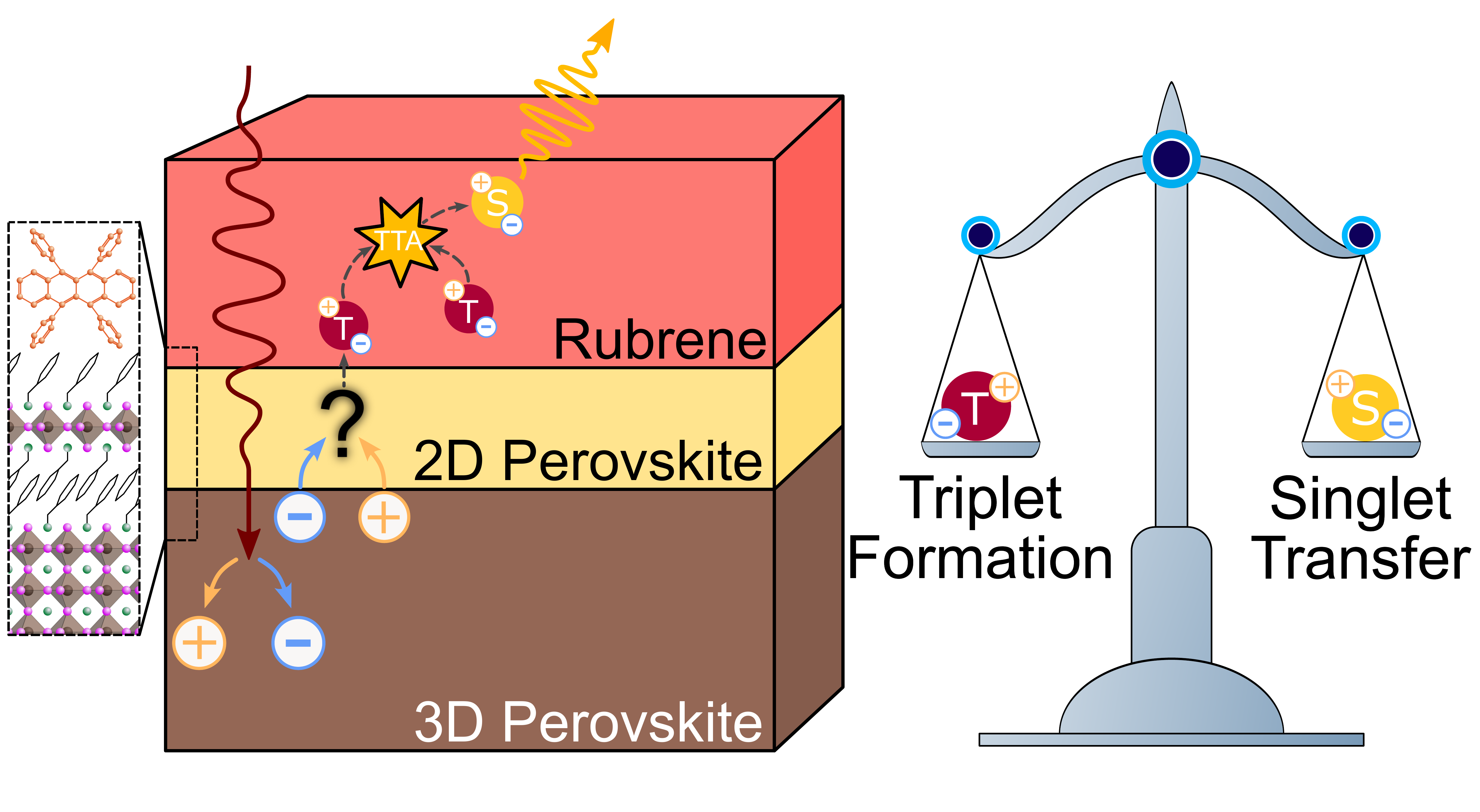}

\end{tocentry}

\begin{abstract}
    Photon upconversion has potential applications in light-emitting diodes, photocatalysis, bio-imaging, microscopy, 3D printing, and photovoltaics. Bulk lead-halide perovskite films have emerged as promising sensitisers for solid-state photon upconversion via triplet-triplet annihilation due to their excellent optoelectronic properties. In this system, a perovskite sensitiser absorbs photons and subsequently generates triplet excitons in an adjacent emitter material, where triplet-triplet annihilation can occur allowing for the emission of higher energy photons. However, a major loss pathway in perovskite-sensitised upconversion is the back-transfer of singlet excitons from the emitter to the sensitiser via F{\"o}rster Resonance Energy Transfer. In this investigation we introduce a 2D perovskite spacer layer between the bulk perovskite sensitiser and a rubrene emitter to mitigate back-transfer of singlet excitons from rubrene to the bulk perovskite sensitiser. This modification reveals the inherent balance between efficient triplet exciton transfer across the interface with a potential barrier versus the mitigation of near-field back-transfer by increasing the distance between the sensitiser and singlet excitons in the emitter. Notably, the introduction of this spacer layer enhances the relative upconversion efficiency at lower excitation power densities while also sustaining performance over extended timescales. This work represents significant progress toward the practical applications of perovskite-sensitised photon upconversion.
\end{abstract}

\section{Introduction}

Photon upconversion is a process where two low-energy photons are converted into one higher-energy photon. This phenomenon shows promise in applications for bio-imaging,\cite{Wang2010UpconversionTherapy,Xu2017HighlyBioimaging} optical microscopy,\cite{Gargas2014EngineeringImaging,Liu2017AmplifiedNanoscopy} photocatalysis,\cite{Khnayzer2011Upconversion-poweredPhotoelectrochemistry,Ravetz2019PhotoredoxUpconversion} 3D printing,\cite{Sanders2022TripletPrinting} reducing the operational voltage for light-emitting diodes,\cite{Pandey2007Rubrene/FullerenePhotovoltage,Chen2016DeterminingDevices,Izawa2023BlueV} and for increasing the efficiency of single junction photovoltaic devices. \cite{Cheng2012ImprovingUpconversion,Schulze2015PhotochemicalConversion,Tayebjee2015BeyondPhotovoltaics,Richards2021PhotonACriticalReview,Naimovicius2023TriplettripletSystems} One mechanism for achieving photon upconversion is triplet-triplet annihilation (TTA), a spin-conserving process in organic semiconductors where two triplet (spin-1) excitons interact to form one higher energy singlet (spin-0) exciton. TTA is particularly promising for  photon upconversion applications, as it can be efficient at sub-solar fluences and with incoherent excitation, as the energy is stored in long-lived triplet excitons.\cite{Singh-Rachford2010PhotonAnnihilation,Schmidt2014PhotochemicalKinetics,Schulze2015PhotochemicalConversion,Sharma2022ConstraintsCells,Feng2023PhotochemicalUpconversion}.

\begin{figure}
    \centering
    \includegraphics[width=\textwidth]{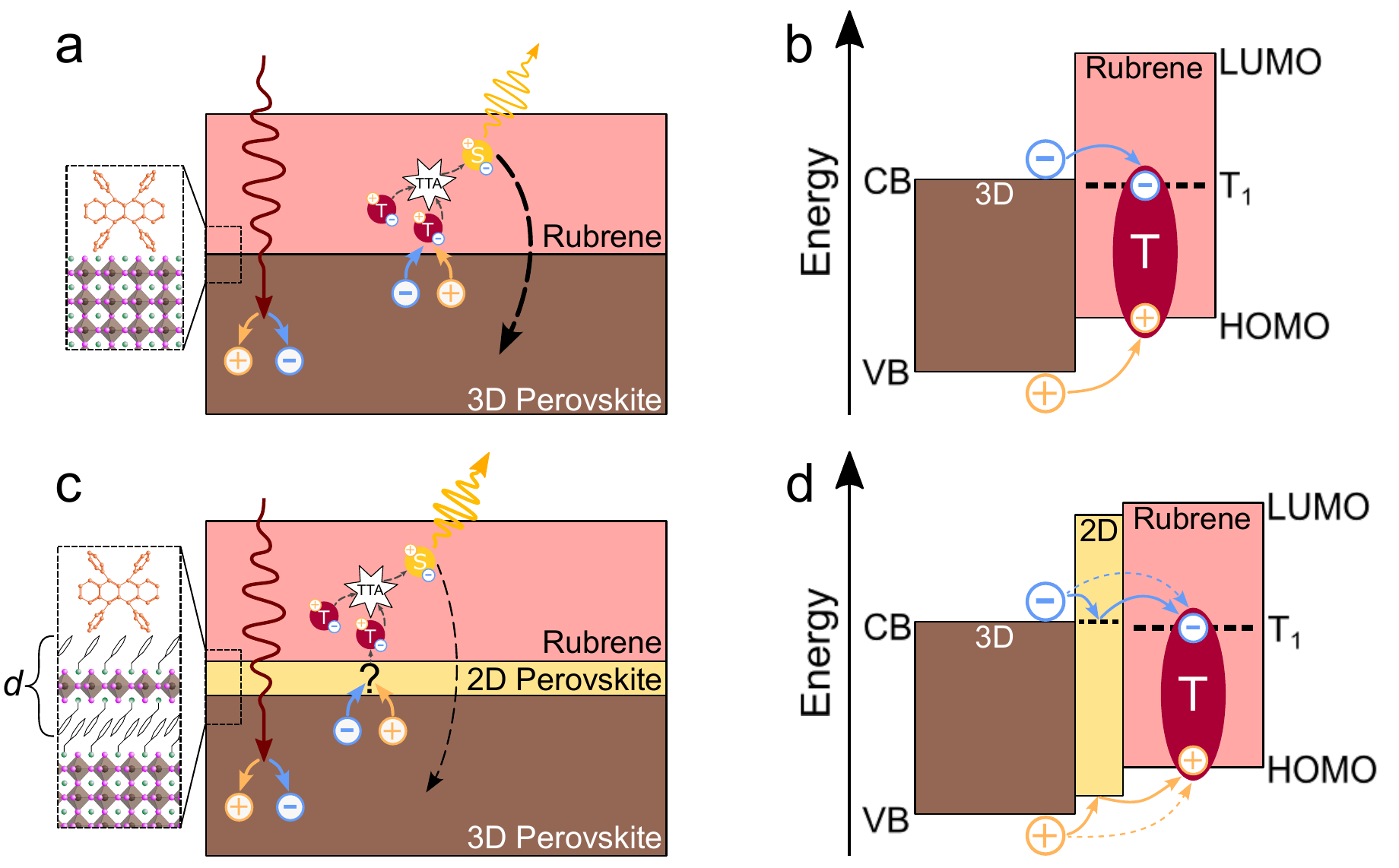}
    \caption{(a) Schematic of the upconversion process in a 3D perovskite/rubrene bilayer. Near-infrared photons are absorbed by the 3D perovskite sensitiser and the photogenerated charge carriers transfer into triplet excitons in the rubrene emitter. These triplet excitons then undergo TTA to form a singlet excitons capable of either photon emission or back-transfer to the perovskite-sensitiser, represented by the dashed arrow. (b) Energetic alignment of a 3D perovskite/rubrene bilayer. The alignment between the perovskite valence band (VB) and the rubrene highest occupied molecular orbital (HOMO), allows for hole transfer. A large energy gap between the perovskite conduction band (CB) and the rubrene lowest unoccupied molecular orbital (LUMO) requires electrons to instead transfer directly into a triplet exciton (T$_1$). (c) Schematic of the upconversion process in the 3D perovskite/2D perovskite/rubrene trilayers investigated in this work. Both electrons and holes transfer through the 2D perovskite spacer layer, resulting in triplet formation and subsequent TTA in the rubrene, however, with anticipated reduced FRET indicated by a thinner dashed arrow. (d) The predicted band alignment for the trilayer system indicates that electrons/holes may transfer through the 2D perovskite spacer layer via direct charge transfer, hopping through defect states, or tunnelling through the 2D perovskite into rubrene.}
    \label{fig:schematic}
\end{figure}

To prevent re-emission of absorbed photons and drive the upconversion process, a ratchet mechanism is required. Ratchet mechanisms are driven by a small sacrifice in free energy to a state that cannot return to the ground state, either due to spatial separation of charges,\cite{Chen2017ANanoribbons} or a change in spin.\cite{Cheng2010KineticLimit,Pusch2019VoltageCells} Here we effect a spin-flip using a sensitiser material coupled to an emitter.\cite{Dexter1953ASolids} The role of the sensitiser is to absorb light and transfer the energy to generate triplet excitons in the emitter material, where TTA can occur. In the case of bulk lead-halide perovskite sensitisers, triplet excitons are populated in the adjacent emitter layer via a sequential charge transfer mechanism,\cite{Nienhaus2019Triplet-SensitizationUpconversion,Wieghold2019TripletFluxes,Sullivan2025AcrossAnnihilation} rather than relying on direct triplet energy transfer from the sensitiser to the emitter.\cite{Singh-Rachford2010PhotonAnnihilation,Huang2015HybridNear-Infrared,Wu2016Solid-stateNanocrystals,Mase2017TripletUpconversion,Chakkamalayath2024EnergyTransfer,Chakkamalayath2024DemystifyingSystem} Following the photoexcitation of free charges in the perovskite sensitiser, holes can transfer from the valence band to the highest molecular orbital (HOMO) of the emitter. In contrast, electrons cannot directly transfer from the conduction band to the lowest unoccupied molecular orbital (LUMO) of the emitter due to a large energy barrier. Instead, electrons can only transfer to the emitter following hole transfer, where they form triplet excitons.\cite{Nienhaus2019Triplet-SensitizationUpconversion,Wieghold2019TripletFluxes} A schematic and a band alignment diagram outlining the upconversion process are shown in \textbf{Figure \ref{fig:schematic}a,b}, respectively. For bulk perovskite sensitised upconverting systems, various emitter molecules have been utilised to maximise the anti-Stokes shift from the upconversion process,\cite{Sullivan2022RechargingReplacement,Sullivan2024TurningCoupling,Sullivan2024WhichAnnihilation} however, rubrene remains the most investigated.\cite{Nienhaus2019Triplet-SensitizationUpconversion,Wieghold2019TripletFluxes,Wieghold2019InfluenceUpconversion,Wieghold2020PrechargingDevices,Wieghold2020IsRubrene,Wieghold2020One-StepDevices,Bieber2020Perovskite-sensitizedTemperature,Prashanthan2020InterdependenceAnnihilators,VanOrman2021EfficiencyMatters,Bieber2021MixedPhonons,Wang2021InterfacialUpconversion,Conti2022UltrafastInterface,Prashanthan2023InternalUpconverters,Sullivan2023SurfaceUpconversion} A drawback of using rubrene as an emitter is that the process of singlet fission is also possible, where a singlet exciton on one chromophore splits to form two triplet excitons on neighbouring chromophores. To counteract singlet fission, the rubrene layer is typically doped with DBP (dibenzotetraphenylperiflanthene).\cite{Okumoto2006HighLayer,Wu2016Solid-stateNanocrystals} The DBP  purportedly harvests singlet excitons from the rubrene via F{\"o}rster Resonance Energy Transfer (FRET) before singlet fission can occur,\cite{Wu2016Solid-stateNanocrystals} however the exact role of DBP in preventing singlet fission lacks consensus.\cite{Wieghold2020IsRubrene,Bossanyi2022InFission}

Lead-halide perovskites are appealing as solid-state sensitisers by virtue of their strong optical absorption, \cite{DeWolf2014OrganometallicPerformance,Bailey2019High-EnergySpectrophotometry} excellent charge transport properties,\cite{Ponseca2014OrganometalRecombination,Galkowski2016DeterminationSemiconductors} and compatibility with solution processing. However, back-transfer of excitons has been a major drawback in their effectiveness in upconverting systems. Singlet excitons generated via TTA in the emitter material, if near the interface, will experience strong near-field FRET back into the perovskite sensitiser due to the large spectral overlap between the singlet exciton emission and the absorption of the perovskite.\cite{Wieghold2019TripletFluxes,Wieghold2019InfluenceUpconversion} Parasitic back-transfer is thus a major loss mechanism in most sensitised upconverting systems and to maximise efficiency, it can be managed by introducing a triplet transmitter layer.\cite{Alves2022ChallengesUpconversion} The role of this layer is to increase the distance between the emitter and sensitizer, thereby reducing the strength of near-field FRET, which decreases proportionally to $1/r^6$ where $r$ is the distance. The thickness of this layer needs to be controlled so that it is sufficiently thick to make FRET negligible while still facilitating effective triplet transfer.\cite{Alves2022ChallengesUpconversion} While triplet transmitting/buffer layers have been utilised for both entirely-organic and quantum dot solid-state upconversion,\cite{Lin2020StrategiesUpconversion,Narayanan2024OvercomingUpconversion} they have yet to be implemented for bulk-perovskite sensitised upconverting systems. 

In this investigation, we introduced thin 2D perovskite films between a bulk perovskite sensitiser and a rubrene emitter to act as both a passivation and spacer layer. 2D perovskite passivation layers have been shown to reduce interfacial recombination, eliminate surface defects, increase moisture resistance, and increase the overall stability of the active layer.\cite{Wu2022SurfaceCells}. We find that the 2D perovskite layers serve a dual purpose: passivating the underlying 3D perovskite sensitizer and improving upconversion efficiency at lower excitation powers by reducing FRET. However, our results also highlight a delicate balance that comes with implementing 2D perovskite spacer layers for TTA upconversion, where increased surface passivation and reduced FRET comes at the expense of decreased triplet sensitisation efficiency. 

\section{Results and Discussion}

To determine the impact of a 2D perovskite spacer layer in a 3D perovskite/rubrene upconverting system, three different spacer layer thicknesses are investigated alongside a control with no 2D perovskite passivation layer. We chose the triple cation, mixed halide composition (Cs$_{0.05}$FA$_{0.79}$MA$_{0.16}$Pb(I$_{0.83}$Br$_{0.17}$)$_3$) subsequently referred to as Br17, for the bulk perovskite sensitiser due to its excellent optical properties.\cite{Saliba2016Cesium-containingEfficiency} The thickness of the 2D perovskite spacer layer is controlled by using three different concentrations of phenethylammonium iodide (PEAI) dissolved in IPA; 1\,mg/mL, 4\,mg/mL, and 8\,mg/mL. For brevity, these samples are subsequently labelled as P1, P4, and P8. The deposited PEAI reacts with unreacted PbI$_2$ precursors on the surface of the Br17 film, forming the Ruddlesden–Popper phase 2D perovskite, PEA$_2$PbI$_4$.\cite{Chen2018} Previously it was demonstrated that an 8 mg/mL PEAI treatment produces a PEA$_2$PbI$_4$ film with an approximate thickness of 30\,nm.\cite{Chen2018} The X-ray diffraction (XRD) and scanning electron microscopy of the samples are shown in \textbf{Figure S1}, and the optical properties (UV-Visible absorption and photoluminescence) are shown in \textbf{Figure S2} and \textbf{S3}, confirming the formation of the PEA$_2$PbI$_4$ spacer layer. PEA$_2$PbI$_4$ was chosen to ensure adequate spectral separation between its absorption onset and the emission of the emitter layer (\textbf{Figure S4}). This emitter layer consists of rubrene doped with 1\% DBP, which is present in all samples, resulting in four upconverting systems; a control sample similar to previously investigated 3D perovskite/rubrene bilayers,\cite{Nienhaus2019Triplet-SensitizationUpconversion,Wieghold2019TripletFluxes,Wieghold2019InfluenceUpconversion,Wieghold2020PrechargingDevices,Wieghold2020IsRubrene,Wieghold2020One-StepDevices,Bieber2020Perovskite-sensitizedTemperature,Prashanthan2020InterdependenceAnnihilators,VanOrman2021EfficiencyMatters,Bieber2021MixedPhonons,Wang2021InterfacialUpconversion,Conti2022UltrafastInterface,Prashanthan2023InternalUpconverters,Sullivan2023SurfaceUpconversion} and three trilayers consisting of a 3D perovskite sensitiser, a 2D perovskite spacer of varying thicknesses, and a rubrene:DBP emitter. These samples are subsequently labelled as Control/Rub, P1/Rub, P4/Rub, and P8/Rub. 

\begin{figure}[!ht]
    \centering
    \includegraphics[width = \linewidth]{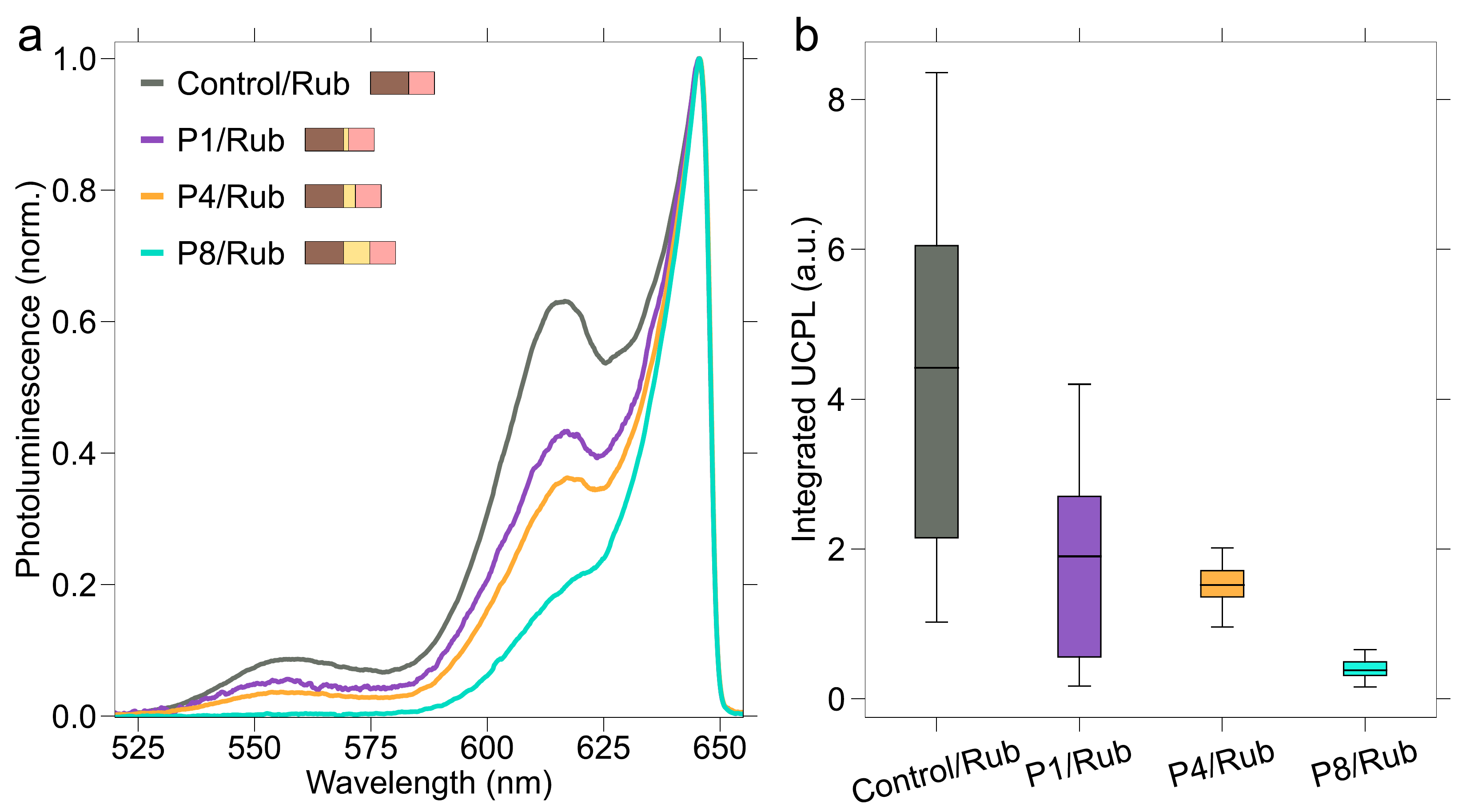}
    \caption{(a) Photoluminescence from the upconverting samples excited at $\lambda_{ex} = 670$\,nm. The cartoons adjacent to the legend labels provide a visual representation of the differences between the samples being compared. (b) Box plot comparing the distribution of the upconverted emission intensity across 4 films for each sample and 5 regions on each film. A 650\,nm shortpass filter was used for collection.}
    \label{fig:ucpl}
\end{figure}

The role of the 3D perovskite film is to absorb incoming low-energy light to generate free charges, which are subsequently transported through the 2D perovskite layer to sensitise triplet excitons in the rubrene film. \textbf{Figure \ref{fig:ucpl}a} shows the upconverted photoluminescence of the samples, excited by a laser with 670 nm wavelength ($\lambda_{ex} =$ 670\,nm). We see that for all samples, upconverted emission is observed from the rubrene:DBP emitter layer ($\mathord{\sim}$520-620\,nm) with varying intensities. Of note is the overlapping signal from the rubrene:DBP emission with the tail of the Br17 emission, so care must be taken when assigning emission to upconversion from the rubrene:DBP layer (see \textbf{Figure S5}). Surprisingly, upconversion is present in the sample with intermediate thickness (P4/Rub) and the thickest capping layer (P8/Rub), however the relative upconversion intensity is very weak for the P8/Rub trilayer sample. To compare the distribution of relative upconversion intensity across all samples, four different films were made for each sample type, and upconversion was measured at five different spots on each film at an excitation power of $\approx 4$\,W/cm$^2$, which we compare in \textbf{Figure \ref{fig:ucpl}b}. Immediately a clear trend can be observed, where increasing spacer layer thickness decreases the intensity of the upconverted emission. This trend is expected, as the introduction of a spacer layer creates an additional potential barrier for charge transfer into the rubrene:DBP emitter layer, reducing the effectiveness of the system to form triplet excitons. Hole transfer through the PEA$_2$PbI$_4$ spacer layer is anticipated, as it is typically utilised in solar cells as a passivation layer between the perovskite active layer and the hole transporting layer.\cite{Chen2018,Cho2018SelectivePhotovoltaics,Jang2021IntactGrowth,Sutanto20212D/3DCells,Teale2024MolecularCells} Nevertheless, electron transfer is surprising as the bandgap of PEA$_2$PbI$_4$ is approximately 0.7\,eV greater than that of Br17. This result raises questions about the charge transfer mechanism, with potential candidates being charges hopping through defect states or tunnelling through the 2D perovskite spacer layer.

\begin{figure}[H]
    \centering
    \includegraphics[width = \linewidth]{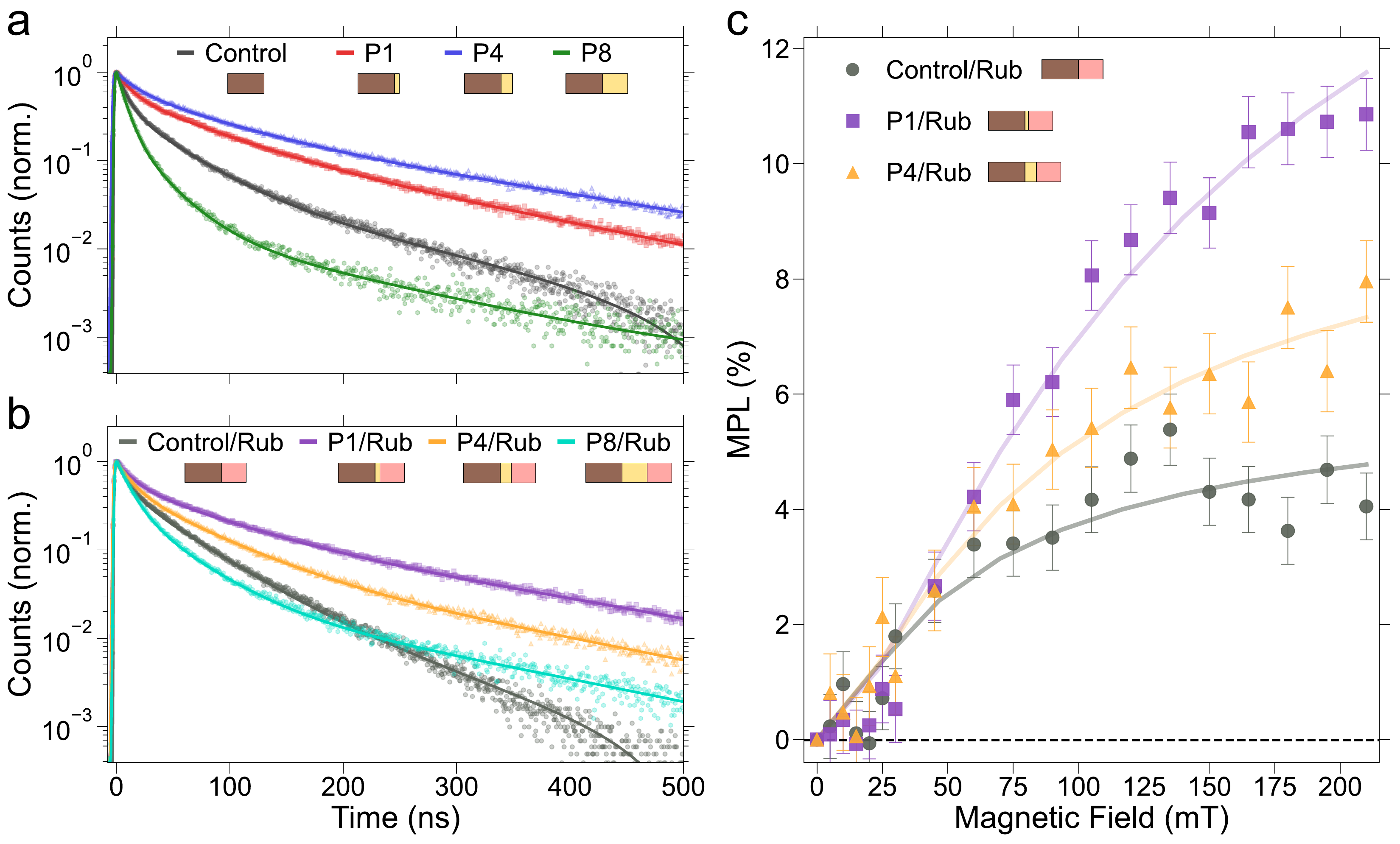}
    \caption{(a) Perovskite PL decay dynamics ($> 725$\,nm) for the (a) neat perovskite and (b) upconverting samples, excited at 700\,nm. The solid lines are triexponential fits with the parameters shown in (a) \textbf{Table S1} and (b) \textbf{S2}. (c) MPL of the upconverted emission ($\lambda < 650$\,nm) from the upconverting samples Control/Rub, P1/Rub, and P4/Rub. The solid lines are approximate fits of a non-Lorentzian function to guide the eye (see Supporting Note 4). The MPL of P8/Rub is omitted as the upconverted emission was too weak to reliably fit via the spectrum decomposition method outlined in \textbf{Figure S5}.}
    \label{fig:TRPL+MPL}
\end{figure}

To investigate the nature of charge transfer through the 2D perovskite layer, we examine the time-resolved photoluminescence from the Br17 layer for both the neat perovskite films and the bi/trilayer samples, shown in \textbf{Figure \ref{fig:TRPL+MPL}a,b} respectively. A triexponential function was fit to the decay traces (see Supporting Note 3), with the averaged lifetimes for both neat and bi/trilayer samples shown in \textbf{Table \ref{table:trpl}}. First, investigating the neat perovskite films, the average lifetime $\bar{\tau}$ of the Control film is calculated to be 31.0 ns. The average lifetime is prolonged for P1 and P4 to 61.6 and 91.7 ns, respectively. The radiative lifetime increase is owed to reduced non-radiative recombination by the passivation of surface trap states in the 3D perovskite.\cite{Chen2018} The sample with the longest $\bar{\tau}$ was found to be P4, demonstrating optimal  passivation of the bulk perovskite surface. Next, comparing the emission lifetime from the Br17 control sample to the Br17/Rub bilayer (\textbf{Table \ref{table:trpl}}), we note that the average lifetime does not change meaningfully (31.0 ns to 31.8 ns). Instead, a reduction in the average lifetimes is seen for both the P1/Rub and P4/Rub trilayers compared to the bare films, potentially indicating charge transfer to the rubrene:DBP layer. However, care must be taken in assigning changes in PL lifetimes for these complex systems. Without the rubrene layer, P4 exhibits a longer $\bar{\tau}$ than P1. However, when the rubrene layer is added, this trend is reversed and the P1/Rub trilayer shows a longer $\bar{\tau}$ than P4/Rub. The increased lifetime of P1/Rub compared to P4/rub is attributed to the FRET-mediated injection of singlets from the rubrene to 3D perovskite, which acts to prolong the PL lifetime of 3D perovskite. Due to the smaller distance between the 3D perovskite and rubrene in P1 compared to P4, the strength of FRET is greater. Thus, the change in lifetimes of the P1/Rub and P4/Rub trilayer samples are attributed to a combination of: (i) the passivation of the surface by the PEAI addition, represented by the increased lifetime compared to the Control Br17 and (ii) the potential reduction in the strength of FRET through the thicker spacer layer, as seen in the shorter $\bar{\tau}$ of the P4/Rub sample compared to the P1/Rub trilayer. For the thickest capping layer (P8), there is evidence of strong quenching of the photoluminescence lifetime, resulting in a reduced $\bar{\tau}$ of 13.2 ns. The P8/Rub trilayer dynamics appear similar to that of P8, likely due to the small amount of charge transfer to the rubrene (evidenced by the overall low UCPL intensity in \textbf{Figure \ref{fig:ucpl}}). Instead, any difference in PL lifetime between P8 and P8/Rub could be due to a complex mix of energy states from the 2D layer itself with \textbf{Figure S3d} showing the range of emissive species present within the thickest 2D PEA$_2$PbI$_4$ layer. One explanation for the existence of these species could be due to the spacer layer consisting of a mixture of perovskite compositions or the presence of higher-order phases (i.e., $\langle n \rangle>1$/quasi-2D phases).\cite{Fu2018MulticolorTransfer,Wen2023HeterojunctionCells} As a result, the thickest spacer layer consists of a complex energetic landscape, which is likely to significantly influence charge transfer through it. The weak upconverted emission and absence of notable interactions in the transient dynamics, as observed in \textbf{Figure \ref{fig:TRPL+MPL}}, further indicates that the 2D perovskite spacer layer prepared with an 8\,mg/mL solution is too thick to allow for effective triplet exciton formation. The increased thickness of the spacer layer is anticipated to significantly hinder charge transfer from the perovskite to rubrene, particularly if the process is reliant on tunnelling.

\begin{table*}[htbp]
  \centering
  \caption{Photoluminescence lifetimes of perovskite emission ($\lambda > 725$\,nm) extracted via triexponential fitting for the samples with and without the rubrene:DBP emitter layer.}
    \begin{tabular}{|>{\columncolor[gray]{0.9}}c|>{\columncolor[gray]{0.95}}c|>{\columncolor[gray]{0.9}}c|}
        \hline
        \textbf{Sample} & $\bar{\tau}$ no Rub:DBP (ns) & $\bar{\tau}$ with Rub:DBP (ns) \\
        \hline
        Control & 31.0 & 31.8 \\
        P1 & 61.6 & 47.1 \\
        P4 & 91.7 & 45.7 \\
        P8 & 13.2 & 25.5 \\
        \hline
    \end{tabular}
  \label{table:trpl}
\end{table*}

To further investigate the upconversion process in the samples, we explore the effects of an externally applied magnetic field on the upconverted light. An externally applied magnetic field is known to influence the dynamics of TTA, described by \citeauthor{Merrifield1968TheoryExcitons} et al.\cite{Merrifield1968TheoryExcitons,Johnson1970EffectsCrystals,Merrifield1971MagneticInteractions} The process of TTA is described in \textit{Equation \ref{eq:TTA}},\cite{Merrifield1971MagneticInteractions} where two independent triplet excitons ($T_1$) combine to form a coupled triplet pair $(TT)$. 

\begin{equation}
    T_1 + T_1 \rightleftharpoons (TT)\rightleftharpoons S_1 + S_0
    \label{eq:TTA}
\end{equation}

The triplet pair state $(TT)$ can occupy nine different spin states, and with no applied field, three of these have singlet character (i.e. with spin-0). With applied fields where the Zeeman splitting is similar in magnitude to the zero-field splitting,\cite{Merrifield1968TheoryExcitons} the number of states with singlet character increases to six and at greater fields decreases to two. This leads to a unique line shape in the magneto-photoluminescence for TTA with a small increase at low applied fields and a decrease at higher fields as only triplet pairs with singlet character can form singlet excitons  to ensure the conservation of spin.

However, TTA is not the only pathway for triplet excitons in the rubrene:DBP layer, since an additional loss mechanism can occur through triplet-charge annihilation (TCA).\cite{Merrifield1971MagneticInteractions,Ern1968MagneticCrystals,Swenburg1973OrganicPhotophysics,Shao2013TripletchargeSemiconductorsb,Thompson2014NanostructuredAnnihilation} The process of TCA is a non-radiative decay pathway for triplet excitons, in which a triplet exciton interacts with a free charge to form a triplet-charge pair $(TC^\pm)$. This interaction can destroy the triplet exciton and generate a single free charge, as described in \textit{Equation \ref{eq:TCA}}.\cite{Merrifield1971MagneticInteractions} 
\begin{equation}
   T_1 + C^\pm \rightleftharpoons {}^\frac{1}{2}(TC^\pm) \rightarrow S_0 + C^\pm 
   \label{eq:TCA}
\end{equation}

At zero field, a triplet-charge pair can occupy six spin states with a mixture of doublet-quartet character. With increasing field the number of spin states with doublet character reduces to four.\cite{Merrifield1971MagneticInteractions, Swenburg1973OrganicPhotophysics,Thompson2014NanostructuredAnnihilation} Due to the conservation of spin, only triplet-charge pairs with doublet character can form a free charge (denoted by ${}^\frac{1}{2}(TC^\pm)$). Thus, with increasing applied field, the rate of TCA decreases, causing the emission from TTA-generated singlet excitons to increase monotonically.

To determine which mechanisms are present within the upconverting samples, we measure the response of the upconverted photoluminescence to an applied magnetic field via magneto-photoluminescence (MPL) defined as:

\begin{equation}
    \text{MPL}(B) \% = \frac{\text{PL}(B) - \text{PL}(0)}{\text{PL}(0)} \times 100
\end{equation}

Where PL(\textit{B}) is the measured photoluminescence intensity at field strength  \textit{B}. Examining \textbf{Figure \ref{fig:TRPL+MPL}c}, we see that for all upconverting samples, with the exception of P8/Rub due to the upconverted emission being too weak, the upconverted emission increases monotonically with the applied field. This result implies that TCA is present and dominant in all systems, pointing towards an inherent charge imbalance, leading to excess free charges in the rubrene layer. These excess charges provide a non-TTA recombination pathway for triplet excitons. Comparing the intensities of the MPL, it is clear that the relative MPL is larger for both P1/Rub and P4/Rub trilayers relative to the Br17 bilayer. This result indicates that the presence of the 2D perovskite spacer layer reduces the effectiveness of free charges to form bound triplet excitons. Based on the predicted energetic alignment at the perovskite/rubrene interface\cite{Ji2017InterfacialInterface,Sloane2025ElectronicTermination} and the energetic alignment between 3D perovskites and 2D passivation layers being typically favourable for hole transfer,\cite{Cho2018SelectivePhotovoltaics,Jang2021IntactGrowth,Sutanto20212D/3DCells,Teale2024MolecularCells,Tang202420.1Perovskites} it is anticipated that holes are more effectively transferred to the rubrene compared to electrons through the passivation layer. It should be noted that while 2D perovskites also exhibit MPL, this typically occurs only at low temperatures and is therefore not expected to influence the observed MPL from the emission of the rubrene:DBP layer.\cite{Bailey2024InfluencePerovskites,Bailey2025RevealingMicroscopy}

For TTA there are typically two regimes seen in the excitation-intensity dependence of the upconverted emission, which is a result of the concentration of triplet excitons in the annihilator.\cite{Cheng2009OnUpconversion,Haefele2012GettingLinear} At low incident power densities, triplet excitons recombine predominantly through first-order non-radiative processes, and the emission from TTA is quadratic with excitation power. However, at higher triplet exciton concentrations, TTA becomes the dominant recombination pathway for triplet excitons, making TTA efficient, resulting in a linear dependence with excitation power. The crossover point from the quadratic dependence (weak regime) to the linear dependence (efficient regime) on excitation power is defined as the threshold intensity $I_{th}$.\cite{Haefele2012GettingLinear}

\begin{figure}
    \centering
    \includegraphics[width = \linewidth]{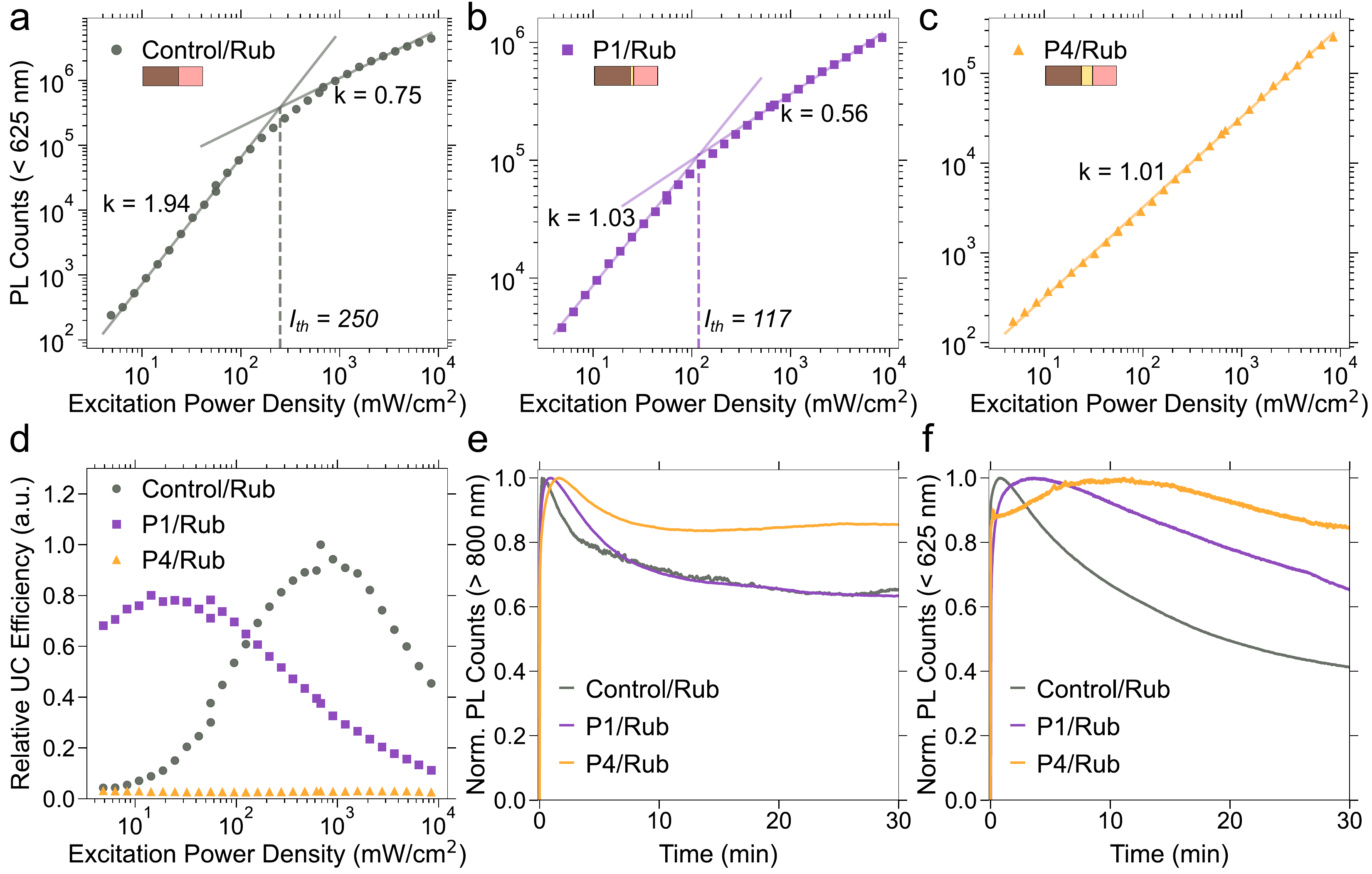}
    \caption{Excitation dependence of the upconverted emission ($\lambda < 625$\,nm) for (a) Control/Rub bilayer, (b) P1/Rub trilayer, and (c) P4/Rub trilayer. Linear fits and the extracted gradients are overlaid, additionally where there is a regime change the $I_{th}$ value is denoted. (d) Relative efficiency of the three upconverting samples defined by upconversion counts/excitation power. (e) Normalised photoluminescence of the bulk perovskite for the upconverting samples over 30 minutes. (f) Normalised upconverted photoluminescence over 30 minutes for the upconverting samples.}
    \label{fig:pdepuc}
\end{figure}

The excitation dependence of the upconverting samples can be  represented by $I_{UC} \propto I_{exc}^k$, where $I_{UC}$ is the upconverted emission intensity, $I_{exc}$ is the excitation intensity, and $k$ is an exponent which is dependent on the recombination mechanism. This relationship is presented in the log-log plots shown in \textbf{Figure \ref{fig:pdepuc}a-c}, where linear fits are applied to distinct regions, and the corresponding slopes yield the values of $k$. The Control/rub bilayer (\textbf{Figure \ref{fig:pdepuc}a}) demonstrates a clear change in the gradient $k$ from 1.94 at low powers to 0.75 at higher powers with a transition at $I_{th} = 250$\,mW/cm$^2$.
The transition to a gradient of $k = 0.75$ is interpreted to be the transition from the weak to efficient annihilation regime, with the deviation of the gradient from the expected value of 1 for efficient TTA due to another carrier-dependent process limiting the efficiency of upconversion, similar to that seen by \citeauthor{Prashanthan2020InterdependenceAnnihilators}\cite{Prashanthan2020InterdependenceAnnihilators} \textbf{Figure \ref{fig:TRPL+MPL}c} shows that all upconverting samples show evidence of TCA, serving as a loss pathway for triplet excitons in the rubrene:DBP layer, which is also affected by the change in excitation power. Correspondingly, for the P1/Rub trilayer (\textbf{Figure \ref{fig:pdepuc}b}) a gradient transition from $k = 1.03$ to $k = 0.56$ is found at $I_{th} = 117$\,mW/cm$^2$. This is an unanticipated result for two main reasons: (i) as seen in \textbf{Figure \ref{fig:ucpl}} the overall upconversion intensity of the P1/Rub trilayer is less than that of the Control/rub bilayer at greater power densities ($\sim$4\,W/cm$^2$) and (ii) although the value of $k$ halves at $I_{th}$, instead of the expected change from $2\rightarrow1$, it is instead $\mathord{\sim}1\rightarrow\mathord{\sim}0.5$. One rationalisation for this observation is the increased presence of TCA in the P1/Rub trilayer (\textbf{Figure \ref{fig:TRPL+MPL}c}): as more charges are photogenerated with increasing excitation power, more of one type of charge carrier will transfer to the rubrene, leading to greater losses of triplet excitons through TCA and thus reduce the value of $k$. Further, the reduction in the value of $I_{th}$ of the P1/Rub trilayer by half compared to the Control/Rub bilayer suggests that the process of TTA in the trilayer is more efficient at lower powers. For the power dependence of TTA to be measured, the upconverted photons must be emitted from the rubrene:DBP layer. We propose that with the presence of the 2D spacer layer, the parasitic back-transfer of singlet excitons is reduced, meaning that a lower concentration of triplet excitons is needed for TTA to become efficient in the P1/Rub trilayer. Comparing these results to the P4/Rub trilayer, no change in gradient is observed; instead, only one gradient of $k = 1.01$ is measured. As the 2D perovskite interlayer increases, there is a decreased driving force for both electrons and holes to transfer into the rubrene, reducing triplet concentration and the rate of TCA, meaning the regime of efficient TTA will not be reached. 

Further investigating the response of the upconverted emission to the excitation power, we compare the Control/Rub bilayer and the two PEAI-treated trilayers by taking the upconverted emission intensity and dividing it by the excitation power (counts/mWcm$^{-2}$), giving a relative upconversion efficiency curve (\textbf{Figure \ref{fig:pdepuc}d}) similar to previous studies.\cite{Prashanthan2023InternalUpconverters} Interestingly, at excitation power densities below $\mathord{\sim}100$\,mW/cm$^2$, P1/Rub exhibits a higher relative upconversion efficiency than the Control/Rub bilayer. Conversely, at higher powers ($>$ $\mathord{\sim}100$\,mW/cm$^2$), the Control/Rub bilayer begins to exhibit a higher relative upconversion efficiency and higher overall counts (\textbf{Figure \ref{fig:ucpl}b}). Thus, the presence of the 2D perovskite interlayer plays an important role: at low excitation power densities, more upconverted photons are being produced by the P1/Rub trilayer sample, indicating a reduction in parasitic back-transfer of TTA-generated singlet excitons to the 3D perovskite sensitiser via FRET. This improvement is particularly relevant for solar applications of photon upconversion, as the observed enhancement occurs at sub-solar power intensities ($\sim< 100$\,mW/cm$^2$), suggesting that the P1/Rub trilayer is more efficient at harnessing low-intensity light for photon upconversion. As the excitation power increases past the optimum point in the relative upconversion efficiency curves, the efficiency begins to decline due to the enhanced rate of TCA caused by the increased generation of free charges. Notably, for the P1/Rub trilayer sample, this reduction in the relative efficiency occurs at higher power densities due to the more prominent TCA seen from the MPL measurement (as seen in \textbf{Figure \ref{fig:TRPL+MPL}c}). In comparison, the P4/Rub trilayer is shows a low relative upconversion efficiency compared to both the Control/Rub bilayer and P1/Rub trilayer, due to the reduced effectiveness of triplet sensitisation through the thicker 2D spacer layer. 

Finally, investigating the relative stability of the emission, both the perovskite PL ($>$ 800\,nm) and the upconversion PL ($<$ 625\,nm) were measured over 30 minutes in \textbf{Figures \ref{fig:pdepuc}e and f} respectively. For the perovskite PL over time (\textbf{Figure \ref{fig:pdepuc}e}), the Control/Rub bilayer shows immediate photo-bleaching following excitation, eventually plateauing at later times ($>$ 10\,min.). A trend is seen for the perovskite PL for the two trilayer samples, showing an increasing photo-brightening time from the thinnest to the intermediate 2D perovskite thicknesses, then followed by a photo-bleach and a subsequent plateau. Reversible photo-bleaching and brightening in 3D perovskites are commonly attributed to ion migration, trap state filling, and changes in non-radiative recombination.\cite{DeQuilettes2016Photo-inducedFilms,Wei2021MechanismsCells} Incorporating 2D perovskite passivation layers can help mitigate these effects by passivating surface defects and suppressing ion movement.\cite{Grancini2017,Wu2022SurfaceCells} Next, we examine how the stability of the upconverted emission changes over time compared to that of the sensitiser. \textbf{Figure \ref{fig:pdepuc}f} shows for the Control/Rub sample that after a short rise time, the upconverted emission continues to decrease over time, reaching $\mathord{\sim}40\%$ of the maximum emission after 30 minutes. The short rise time is interpreted as interfacial trap states becoming saturated following continuous emission, meaning trap-filling no longer competes with charge transfer to the rubrene.\cite{VanOrman2021EfficiencyMatters} However, the photo-bleach likely stems from the same source of the photo-bleaching of the perovskite PL seen in \textbf{Figure \ref{fig:pdepuc}e}. Similar to \textbf{Figure \ref{fig:pdepuc}e}, the two trilayer samples both show increasing photo-brightening rise time with increasing 2D perovskite thickness however, they also show greater stability over the 30-minute measurement window with the trilayers retaining $\mathord{\sim}65\%$ and $\mathord{\sim}85\%$ of their maximum upconverted intensity for the P1/Rub and the P4/Rub trilayers respectively.

The results of this investigation suggest that trilayer systems hold significant promise for upconversion applications. The P1/Rub sample exhibited a lower $I_{th}$ than the Control/Rub, and showed a higher relative upconversion efficiency at lower excitation densities. Additionally, the trilayer samples also exhibit more stable upconversion emission over extended time periods compared to the Control/Rub bilayer. However, there remains significant potential for improvement, namely via utilising different compositions/orientations for the 2D perovskite. It is clear the main drawback of utilising a 2D perovskite spacer layer is the reduction in triplet exciton generation in the rubrene:DBP emitting layer likely due to the poor charge transport properties of 2D perovskites as effective charge transport is confined to the spatially separated octahedral layers. One potential method to address this is using quasi-2D perovskites, which contain multiple octahedral layers and possess better charge transport, or using functionalised spacer cations to enhance charge transfer.\cite{Zhao2022a} Alternatively, using 2D perovskite layers with perpendicularly oriented octahedral layers to the 3D perovskite substrate would also enhance charge transport and has shown promise for application in photovoltaics.\cite{Zhang2017VerticallyStability,Chen2018OriginPerformance} Regardless of the methods employed, two major criteria need to be fulfilled for 2D perovskite spacer layers to further progress in improving solid-state upconversion, (i) charge transfer from the bulk sensitiser to the organic emitter needs to occur through the 2D spacer layer and (ii) the absorption onset of the 2D spacer layer should not overlap with the emission from the emitting layer.\cite{Alves2022ChallengesUpconversion}

\section{Conclusion}

In conclusion, we have investigated the impact of introducing a 2D perovskite layer between a 3D perovskite sensitiser and a rubrene:DBP emitter layer for photon upconversion. A clear benefit is the overall increased stability of the 3D perovskite due to the 2D perovskite layer, which is the main role the 2D perovskite passivation layer has played extensively in the literature. However, we also show that there is benefit in increasing the distance between TTA-generated singlet excitons in the rubrene and the strongly absorbing 3D perovskite. The introduction of a thin PEA$_2$PbI$_4$ layer showed a stronger upconversion intensity at lower excitation powers and more stable upconverted PL. This result demonstrates that for photon upconversion applications, the bulk sensitiser/2D spacer/organic emitter trilayers show considerable promise under constant excitation at sub-solar power densities (i.e. $< 100$\,mW/cm$^2$). However, there are clear trade-offs when introducing the 2D spacer layer, with the overall upconversion intensity decreasing with increasing 2D layer thickness. This is attributed to both increased TCA, arising from a greater charge imbalance in the rubrene:DBP layer, and the larger potential barrier for charge transport from the 3D perovskite to the organic emitter, which is constrained by carrier hopping through intermediate defect states or by tunnelling. A clear balance must be considered: while the 2D perovskite layer enhances stability and prevents parasitic back-transfer, it reduces the efficiency of photon upconversion at higher excitation densities.

\section{Experimental Methods}

\textbf{\textit{Materials}}

Unless stated otherwise, all materials were purchased from Sigma Aldrich. Formamidinium iodide (FAI), phenethylammonium iodide (PEAI), and methylammonium bromide (MABr) were purchased from GreatCell Solar Materials, and Lead iodide (PbI$_2$) was purchased from TCI. Rubrene ($>$\,99\%) and DBP ($>$\,99\%) were purchased from Ossila.

\textbf{\textit{Perovskite Fabrication}}

Glass substrates were sequentially cleaned with Hellmanex™ III cleaning solution, deionised water, acetone and 2-propanol for a duration of 10 minutes for each step using an ultrasonic bath. The substrates were then treated in a UV-ozone machine for 15 minutes. To prepare the perovskite precursor, 172\,mg formamidinium iodide (FAI), 22.4\,mg methylammonium bromide (MABr), 507.1\,mg lead iodide (PbI$_2$), and 73.4\,mg lead bromide (PbBr$_2$) were mixed in an 1 ml solvent mixture of N,N-dimethylformamide (DMF) and dimethyl sulfoxide (DMSO) with a volume ratio of 4:1. Then the precursor was doped with 35\,$\mu$L cesium iodide (CsI) from a stock solution (in DMSO) of 1.4M concentration. A two-step spin-coating program coats the perovskite film on the substrates. After covering the substrate with perovskite precursor solution, the substrate was first rotated at a spin rate of 2000 rpm for 10 s and then at a spin rate of 6000\,rpm for 30\,s. During the second step, $\sim 150$\,$\mu$L of chlorobenzene was dropped on the centre of the spinning substrate 15s prior to the end of the spinning program. Immediately after spin-coating, the substrates were placed on a hotplate and annealed for 10 minutes at 100 $^\circ$C.  For samples with phenethylammonium iodide (PEAI) based capping layers, precursor solutions of different concentrations (1\,mg/ml, 4\,mg/ml, and 8\,mg/ml in 2-propanol) were spin-coated on perovskite coated substrates at a spin rate of 5000 rpm for 15 s. This was followed by an annealing step of 10 minutes at 100$^\circ$C.

\textbf{\textit{Bi/Trilayer Fabrication}}

A rubrene:DBP solution was prepared by first dissolving rubrene in chlorobenzene at 10\,mg/mL and dissolving DBP in chlorobenzene at 1\,mg/mL. The rubrene:DBP solution was made by doping the rubrene with the DBP at a 1\% molar ratio. This solution was then deposited on top of all perovskite films through spin-coating via the dynamic method at 6000\,rpm for 30 seconds. The films were then annealed on a hotplate at 100$^\circ$ C for 10 minutes. To prevent exposure to oxygen/moisture the films where then encapsulated using a glass slide and UV-curing epoxy inside the nitrogen filled glovebox.

\textbf{\textit{X-Ray Diffraction}}

X-ray diffraction patterns were measured by a PANalytical Xpert materials research diffractometer (MRD) with Cu K$\alpha$ radiation ($\lambda$ = 1.54056\,Å), using an accelerating voltage of 45\,kV and a current of 40\,mA.

\textbf{\textit{Scanning Electron Microscopy}}

The SEM images were obtained using a FEI NanoSEM 450 field-emission scanning electron microscope operating at 5\,kV.

\textbf{\textit{UV-Vis Absorption}}

UV–vis absorbance of the thin films was measured using a UV-2600 Shimadzu spectrophotometer with the integrating sphere attachment ISR2600.

\textbf{\textit{Steady-State Photoluminescence}}

All photoluminescence measurements were measured using off-axis parabolic mirrors for collection, and the spectra shown were collected using a Thorlabs CS200 fibre-coupled spectrometer. The spectra in \textbf{Figure S3} were excited using a Thorlabs CPS405 405\,nm diode laser. For the spectra/boxplot in \textbf{Figure \ref{fig:ucpl}} the samples were excited using a BWTEK BWF1 Fibre Coupled 670\,nm laser with an excitation density $\approx 4$W/cm$^2$ and the signal was collected through a 650\,nm short-pass filter
For excitation dependence the samples were excited by a Cobalt 06 MLD-730 fibre coupled laser. To change the excitation power the collimated beam was passed through a linear polariser followed by a half wave plate on a motorised stage followed by another linear polariser and the power was calibrated. Photoluminescence counts were measured by an Excelitas SPCM-AQRH Avalanche Photodiode, the upconverted emission was passed through a 625\,nm short-pass filter and the emission from the perovskite was passed through an 800\,nm long-pass filter.

\textbf{\textit{Time Resolved Photoluminescence}}

Time-correlated single photon counting (TCSPC) was measured with a benchtop microscope. The sample was excited with a 700\,nm picosecond pulsed laser (EKSPLA, PT200, 10 ps pulse-width, 1 MHz). A 0.3NA (15$\times$) silver reflective objective was used to collect emissions from the sample. The excitation beam was filtered out with a 725\,nm long pass filter and the emission was detected with a silicon photodiode detector (IQ Quantique, ID110). The signal was digitised using a TCSPC card (PicoQuant, TimeHarp 260).

\textbf{\textit{Magneto-Photoluminescence}}

The samples were excited with a $670$\,nm laser (Thorlabs CPS670F). A bench top electromagnet (Magnetech MFG-6-24) controlled by a D.C. power supply (Keithley 2230G-30-1) produced the external magnetic field. An Ocean Optics Flame spectrometer collected the photoluminescence spectrum. The field was calibrated with a Gaussmeter (Lakeshore 475).

\begin{acknowledgement}

The authors acknowledge support through the Australian Research Council Centre of Excellence in Exciton Science (CE170100026). N.P.S and D.M.d.C acknowledges the support from an Australian Government Research Training Program (RTP) Scholarship. This work is also supported by the Australian Renewable Energy Agency (ARENA). A.W.Y.H.-B. is supported by the Australian Research Council (ARC) Future Fellowship FT210100210. M.P.N. recognises the support of the UNSW Scientia Program and an ARC DECRA Fellowship (DE230100382). The authors acknowledge the facilities and the scientific and technical assistance of Microscopy Australia at the Electron Microscope Unit (EMU) and the Solid State and Elemental Analysis Unit (SSEAU) within the Mark Wainwright Analytical Centre (MWAC) at UNSW Sydney. 

\end{acknowledgement}
\bibliography{manuscript_references}

\providecommand{\latin}[1]{#1}
\makeatletter
\providecommand{\doi}
  {\begingroup\let\do\@makeother\dospecials
  \catcode`\{=1 \catcode`\}=2 \doi@aux}
\providecommand{\doi@aux}[1]{\endgroup\texttt{#1}}
\makeatother
\providecommand*\mcitethebibliography{\thebibliography}
\csname @ifundefined\endcsname{endmcitethebibliography}  {\let\endmcitethebibliography\endthebibliography}{}
\begin{mcitethebibliography}{85}
\providecommand*\natexlab[1]{#1}
\providecommand*\mciteSetBstSublistMode[1]{}
\providecommand*\mciteSetBstMaxWidthForm[2]{}
\providecommand*\mciteBstWouldAddEndPuncttrue
  {\def\EndOfBibitem{\unskip.}}
\providecommand*\mciteBstWouldAddEndPunctfalse
  {\let\EndOfBibitem\relax}
\providecommand*\mciteSetBstMidEndSepPunct[3]{}
\providecommand*\mciteSetBstSublistLabelBeginEnd[3]{}
\providecommand*\EndOfBibitem{}
\mciteSetBstSublistMode{f}
\mciteSetBstMaxWidthForm{subitem}{(\alph{mcitesubitemcount})}
\mciteSetBstSublistLabelBeginEnd
  {\mcitemaxwidthsubitemform\space}
  {\relax}
  {\relax}

\bibitem[Wang \latin{et~al.}(2010)Wang, Banerjee, Liu, Chen, and Liu]{Wang2010UpconversionTherapy}
Wang,~F.; Banerjee,~D.; Liu,~Y.; Chen,~X.; Liu,~X. {Upconversion nanoparticles in biological labeling, imaging, and therapy}. \emph{Analyst} \textbf{2010}, \emph{135}, 1839--1854\relax
\mciteBstWouldAddEndPuncttrue
\mciteSetBstMidEndSepPunct{\mcitedefaultmidpunct}
{\mcitedefaultendpunct}{\mcitedefaultseppunct}\relax
\EndOfBibitem
\bibitem[Xu \latin{et~al.}(2017)Xu, Yang, Sun, Bi, Liu, Yang, Gai, He, and Lin]{Xu2017HighlyBioimaging}
Xu,~J.; Yang,~P.; Sun,~M.; Bi,~H.; Liu,~B.; Yang,~D.; Gai,~S.; He,~F.; Lin,~J. {Highly Emissive Dye-Sensitized Upconversion Nanostructure for Dual-Photosensitizer Photodynamic Therapy and Bioimaging}. \emph{ACS Nano} \textbf{2017}, \emph{11}, 4133--4144\relax
\mciteBstWouldAddEndPuncttrue
\mciteSetBstMidEndSepPunct{\mcitedefaultmidpunct}
{\mcitedefaultendpunct}{\mcitedefaultseppunct}\relax
\EndOfBibitem
\bibitem[Gargas \latin{et~al.}(2014)Gargas, Chan, Ostrowski, Aloni, Altoe, Barnard, Sanii, Urban, Milliron, Cohen, and Schuck]{Gargas2014EngineeringImaging}
Gargas,~D.~J.; Chan,~E.~M.; Ostrowski,~A.~D.; Aloni,~S.; Altoe,~M. V.~P.; Barnard,~E.~S.; Sanii,~B.; Urban,~J.~J.; Milliron,~D.~J.; Cohen,~B.~E.; Schuck,~P.~J. {Engineering bright sub-10-nm upconverting nanocrystals for single-molecule imaging}. \emph{Nature Nanotechnology} \textbf{2014}, \emph{9}, 300--305\relax
\mciteBstWouldAddEndPuncttrue
\mciteSetBstMidEndSepPunct{\mcitedefaultmidpunct}
{\mcitedefaultendpunct}{\mcitedefaultseppunct}\relax
\EndOfBibitem
\bibitem[Liu \latin{et~al.}(2017)Liu, Lu, Yang, Zheng, Wen, Wang, Vidal, Zhao, Liu, Zhou, Ma, Zhou, Piper, Xi, and Jin]{Liu2017AmplifiedNanoscopy}
Liu,~Y.; Lu,~Y.; Yang,~X.; Zheng,~X.; Wen,~S.; Wang,~F.; Vidal,~X.; Zhao,~J.; Liu,~D.; Zhou,~Z.; Ma,~C.; Zhou,~J.; Piper,~J.~A.; Xi,~P.; Jin,~D. {Amplified stimulated emission in upconversion nanoparticles for super-resolution nanoscopy}. \emph{Nature} \textbf{2017}, \emph{543}, 229--233\relax
\mciteBstWouldAddEndPuncttrue
\mciteSetBstMidEndSepPunct{\mcitedefaultmidpunct}
{\mcitedefaultendpunct}{\mcitedefaultseppunct}\relax
\EndOfBibitem
\bibitem[Khnayzer \latin{et~al.}(2011)Khnayzer, Blumhoff, Harrington, Haefele, Deng, and Castellano]{Khnayzer2011Upconversion-poweredPhotoelectrochemistry}
Khnayzer,~R.~S.; Blumhoff,~J.; Harrington,~J.~A.; Haefele,~A.; Deng,~F.; Castellano,~F.~N. {Upconversion-powered photoelectrochemistry}. \emph{Chemical Communications} \textbf{2011}, \emph{48}, 209--211\relax
\mciteBstWouldAddEndPuncttrue
\mciteSetBstMidEndSepPunct{\mcitedefaultmidpunct}
{\mcitedefaultendpunct}{\mcitedefaultseppunct}\relax
\EndOfBibitem
\bibitem[Ravetz \latin{et~al.}(2019)Ravetz, Pun, Churchill, Congreve, Rovis, and Campos]{Ravetz2019PhotoredoxUpconversion}
Ravetz,~B.~D.; Pun,~A.~B.; Churchill,~E.~M.; Congreve,~D.~N.; Rovis,~T.; Campos,~L.~M. {Photoredox catalysis using infrared light via triplet fusion upconversion}. \emph{Nature} \textbf{2019}, \emph{565}, 343--346\relax
\mciteBstWouldAddEndPuncttrue
\mciteSetBstMidEndSepPunct{\mcitedefaultmidpunct}
{\mcitedefaultendpunct}{\mcitedefaultseppunct}\relax
\EndOfBibitem
\bibitem[Sanders \latin{et~al.}(2022)Sanders, Schloemer, Gangishetty, Anderson, Seitz, Gallegos, Stokes, and Congreve]{Sanders2022TripletPrinting}
Sanders,~S.~N.; Schloemer,~T.~H.; Gangishetty,~M.~K.; Anderson,~D.; Seitz,~M.; Gallegos,~A.~O.; Stokes,~R.~C.; Congreve,~D.~N. {Triplet fusion upconversion nanocapsules for volumetric 3D printing}. \emph{Nature} \textbf{2022}, \emph{604}, 474--478\relax
\mciteBstWouldAddEndPuncttrue
\mciteSetBstMidEndSepPunct{\mcitedefaultmidpunct}
{\mcitedefaultendpunct}{\mcitedefaultseppunct}\relax
\EndOfBibitem
\bibitem[Pandey and Nunzi(2007)Pandey, and Nunzi]{Pandey2007Rubrene/FullerenePhotovoltage}
Pandey,~A.~K.; Nunzi,~J.~M. {Rubrene/Fullerene Heterostructures with a Half-Gap Electroluminescence Threshold and Large Photovoltage}. \emph{Advanced Materials} \textbf{2007}, \emph{19}, 3613--3617\relax
\mciteBstWouldAddEndPuncttrue
\mciteSetBstMidEndSepPunct{\mcitedefaultmidpunct}
{\mcitedefaultendpunct}{\mcitedefaultseppunct}\relax
\EndOfBibitem
\bibitem[Chen \latin{et~al.}(2016)Chen, Jia, Chen, Yuan, Zou, and Xiong]{Chen2016DeterminingDevices}
Chen,~Q.; Jia,~W.; Chen,~L.; Yuan,~D.; Zou,~Y.; Xiong,~Z. {Determining the Origin of Half-bandgap-voltage Electroluminescence in Bifunctional Rubrene/C60 Devices}. \emph{Scientific Reports} \textbf{2016}, \emph{6}, 25331\relax
\mciteBstWouldAddEndPuncttrue
\mciteSetBstMidEndSepPunct{\mcitedefaultmidpunct}
{\mcitedefaultendpunct}{\mcitedefaultseppunct}\relax
\EndOfBibitem
\bibitem[Izawa \latin{et~al.}(2023)Izawa, Morimoto, Fujimoto, Banno, Majima, Takahashi, Naka, and Hiramoto]{Izawa2023BlueV}
Izawa,~S.; Morimoto,~M.; Fujimoto,~K.; Banno,~K.; Majima,~Y.; Takahashi,~M.; Naka,~S.; Hiramoto,~M. {Blue organic light-emitting diode with a turn-on voltage of 1.47 V}. \emph{Nature Communications} \textbf{2023}, \emph{14}, 5494\relax
\mciteBstWouldAddEndPuncttrue
\mciteSetBstMidEndSepPunct{\mcitedefaultmidpunct}
{\mcitedefaultendpunct}{\mcitedefaultseppunct}\relax
\EndOfBibitem
\bibitem[Cheng \latin{et~al.}(2012)Cheng, F{\"{u}}ckel, MacQueen, Khoury, Clady, Schulze, Ekins-Daukes, Crossley, Stannowski, Lips, and Schmidt]{Cheng2012ImprovingUpconversion}
Cheng,~Y.~Y.; F{\"{u}}ckel,~B.; MacQueen,~R.~W.; Khoury,~T.; Clady,~R. G. C.~R.; Schulze,~T.~F.; Ekins-Daukes,~N.~J.; Crossley,~M.~J.; Stannowski,~B.; Lips,~K.; Schmidt,~T.~W. {Improving the light-harvesting of amorphous silicon solar cells with photochemical upconversion}. \emph{Energy {\&} Environmental Science} \textbf{2012}, \emph{5}, 6953\relax
\mciteBstWouldAddEndPuncttrue
\mciteSetBstMidEndSepPunct{\mcitedefaultmidpunct}
{\mcitedefaultendpunct}{\mcitedefaultseppunct}\relax
\EndOfBibitem
\bibitem[Schulze and Schmidt(2015)Schulze, and Schmidt]{Schulze2015PhotochemicalConversion}
Schulze,~T.~F.; Schmidt,~T.~W. {Photochemical upconversion: present status and prospects for its application to solar energy conversion}. \emph{Energy {\&} Environmental Science} \textbf{2015}, \emph{8}, 103--125\relax
\mciteBstWouldAddEndPuncttrue
\mciteSetBstMidEndSepPunct{\mcitedefaultmidpunct}
{\mcitedefaultendpunct}{\mcitedefaultseppunct}\relax
\EndOfBibitem
\bibitem[Tayebjee \latin{et~al.}(2015)Tayebjee, McCamey, and Schmidt]{Tayebjee2015BeyondPhotovoltaics}
Tayebjee,~M. J.~Y.; McCamey,~D.~R.; Schmidt,~T.~W. {Beyond Shockley–Queisser: Molecular Approaches to High-Efficiency Photovoltaics}. \emph{The Journal of Physical Chemistry Letters} \textbf{2015}, \emph{6}, 2367--2378\relax
\mciteBstWouldAddEndPuncttrue
\mciteSetBstMidEndSepPunct{\mcitedefaultmidpunct}
{\mcitedefaultendpunct}{\mcitedefaultseppunct}\relax
\EndOfBibitem
\bibitem[Richards \latin{et~al.}(2021)Richards, Hudry, Busko, Turshatov, and Howard]{Richards2021PhotonACriticalReview}
Richards,~B.~S.; Hudry,~D.; Busko,~D.; Turshatov,~A.; Howard,~I.~A. {Photon Upconversion for Photovoltaics and Photocatalysis: A Critical Review}. \emph{Chemical Reviews} \textbf{2021}, \emph{121}, 9165--9195\relax
\mciteBstWouldAddEndPuncttrue
\mciteSetBstMidEndSepPunct{\mcitedefaultmidpunct}
{\mcitedefaultendpunct}{\mcitedefaultseppunct}\relax
\EndOfBibitem
\bibitem[Naimovi{\v{c}}ius \latin{et~al.}(2023)Naimovi{\v{c}}ius, Bharmoria, and Moth-Poulsen]{Naimovicius2023TriplettripletSystems}
Naimovi{\v{c}}ius,~L.; Bharmoria,~P.; Moth-Poulsen,~K. {Triplet–triplet annihilation mediated photon upconversion solar energy systems}. \emph{Materials Chemistry Frontiers} \textbf{2023}, \emph{7}, 2297--2315\relax
\mciteBstWouldAddEndPuncttrue
\mciteSetBstMidEndSepPunct{\mcitedefaultmidpunct}
{\mcitedefaultendpunct}{\mcitedefaultseppunct}\relax
\EndOfBibitem
\bibitem[Singh-Rachford and Castellano(2010)Singh-Rachford, and Castellano]{Singh-Rachford2010PhotonAnnihilation}
Singh-Rachford,~T.~N.; Castellano,~F.~N. {Photon upconversion based on sensitized triplet–triplet annihilation}. \emph{Coordination Chemistry Reviews} \textbf{2010}, \emph{254}, 2560--2573\relax
\mciteBstWouldAddEndPuncttrue
\mciteSetBstMidEndSepPunct{\mcitedefaultmidpunct}
{\mcitedefaultendpunct}{\mcitedefaultseppunct}\relax
\EndOfBibitem
\bibitem[Schmidt and Castellano(2014)Schmidt, and Castellano]{Schmidt2014PhotochemicalKinetics}
Schmidt,~T.~W.; Castellano,~F.~N. {Photochemical Upconversion: The Primacy of Kinetics}. \emph{The Journal of Physical Chemistry Letters} \textbf{2014}, \emph{5}, 4062--4072\relax
\mciteBstWouldAddEndPuncttrue
\mciteSetBstMidEndSepPunct{\mcitedefaultmidpunct}
{\mcitedefaultendpunct}{\mcitedefaultseppunct}\relax
\EndOfBibitem
\bibitem[Sharma \latin{et~al.}(2022)Sharma, Pusch, Nielsen, R{\"{o}}mer, Tayebjee, Rougieux, and Ekins-Daukes]{Sharma2022ConstraintsCells}
Sharma,~A.~S.; Pusch,~A.; Nielsen,~M.~P.; R{\"{o}}mer,~U.; Tayebjee,~M.~J.; Rougieux,~F.~E.; Ekins-Daukes,~N.~J. {Constraints imposed by the sparse solar photon flux on upconversion and hot carrier solar cells}. \emph{Solar Energy} \textbf{2022}, \emph{237}, 44--51\relax
\mciteBstWouldAddEndPuncttrue
\mciteSetBstMidEndSepPunct{\mcitedefaultmidpunct}
{\mcitedefaultendpunct}{\mcitedefaultseppunct}\relax
\EndOfBibitem
\bibitem[Feng \latin{et~al.}(2023)Feng, Alves, de~Clercq, and Schmidt]{Feng2023PhotochemicalUpconversion}
Feng,~J.; Alves,~J.; de~Clercq,~D.~M.; Schmidt,~T.~W. {Photochemical Upconversion}. \emph{Annual Review of Physical Chemistry} \textbf{2023}, \emph{74}, 145--168\relax
\mciteBstWouldAddEndPuncttrue
\mciteSetBstMidEndSepPunct{\mcitedefaultmidpunct}
{\mcitedefaultendpunct}{\mcitedefaultseppunct}\relax
\EndOfBibitem
\bibitem[Chen and Wu(2017)Chen, and Wu]{Chen2017ANanoribbons}
Chen,~S.-F.; Wu,~Y.-R. {A design of intermediate band solar cell for photon ratchet with multi-layer MoS$_2$ nanoribbons}. \emph{Applied Physics Letters} \textbf{2017}, \emph{110}\relax
\mciteBstWouldAddEndPuncttrue
\mciteSetBstMidEndSepPunct{\mcitedefaultmidpunct}
{\mcitedefaultendpunct}{\mcitedefaultseppunct}\relax
\EndOfBibitem
\bibitem[Cheng \latin{et~al.}(2010)Cheng, F{\"{u}}ckel, Khoury, Clady, Tayebjee, Ekins-Daukes, Crossley, and Schmidt]{Cheng2010KineticLimit}
Cheng,~Y.~Y.; F{\"{u}}ckel,~B.; Khoury,~T.; Clady,~R. G. C.~R.; Tayebjee,~M. J.~Y.; Ekins-Daukes,~N.~J.; Crossley,~M.~J.; Schmidt,~T.~W. {Kinetic Analysis of Photochemical Upconversion by Triplet-Triplet Annihilation: Beyond Any Spin Statistical Limit}. \emph{The Journal of Physical Chemistry Letters} \textbf{2010}, \emph{1}, 1795--1799\relax
\mciteBstWouldAddEndPuncttrue
\mciteSetBstMidEndSepPunct{\mcitedefaultmidpunct}
{\mcitedefaultendpunct}{\mcitedefaultseppunct}\relax
\EndOfBibitem
\bibitem[Pusch and Ekins-Daukes(2019)Pusch, and Ekins-Daukes]{Pusch2019VoltageCells}
Pusch,~A.; Ekins-Daukes,~N.~J. {Voltage Matching, {\'{E}}tendue, and Ratchet Steps in Advanced-Concept Solar Cells}. \emph{Physical Review Applied} \textbf{2019}, \emph{12}, 044055\relax
\mciteBstWouldAddEndPuncttrue
\mciteSetBstMidEndSepPunct{\mcitedefaultmidpunct}
{\mcitedefaultendpunct}{\mcitedefaultseppunct}\relax
\EndOfBibitem
\bibitem[Dexter(1953)]{Dexter1953ASolids}
Dexter,~D.~L. {A Theory of Sensitized Luminescence in Solids}. \emph{The Journal of Chemical Physics} \textbf{1953}, \emph{21}, 836--850\relax
\mciteBstWouldAddEndPuncttrue
\mciteSetBstMidEndSepPunct{\mcitedefaultmidpunct}
{\mcitedefaultendpunct}{\mcitedefaultseppunct}\relax
\EndOfBibitem
\bibitem[Nienhaus \latin{et~al.}(2019)Nienhaus, Correa-Baena, Wieghold, Einzinger, Lin, Shulenberger, Klein, Wu, Bulovi{\'{c}}, Buonassisi, Baldo, and Bawendi]{Nienhaus2019Triplet-SensitizationUpconversion}
Nienhaus,~L.; Correa-Baena,~J.-P.; Wieghold,~S.; Einzinger,~M.; Lin,~T.-A.; Shulenberger,~K.~E.; Klein,~N.~D.; Wu,~M.; Bulovi{\'{c}},~V.; Buonassisi,~T.; Baldo,~M.~A.; Bawendi,~M.~G. {Triplet-Sensitization by Lead Halide Perovskite Thin Films for Near-Infrared-to-Visible Upconversion}. \emph{ACS Energy Letters} \textbf{2019}, \emph{4}, 888--895\relax
\mciteBstWouldAddEndPuncttrue
\mciteSetBstMidEndSepPunct{\mcitedefaultmidpunct}
{\mcitedefaultendpunct}{\mcitedefaultseppunct}\relax
\EndOfBibitem
\bibitem[Wieghold \latin{et~al.}(2019)Wieghold, Bieber, VanOrman, Daley, Leger, Correa-Baena, and Nienhaus]{Wieghold2019TripletFluxes}
Wieghold,~S.; Bieber,~A.~S.; VanOrman,~Z.~A.; Daley,~L.; Leger,~M.; Correa-Baena,~J.~P.; Nienhaus,~L. {Triplet Sensitization by Lead Halide Perovskite Thin Films for Efficient Solid-State Photon Upconversion at Subsolar Fluxes}. \emph{Matter} \textbf{2019}, \emph{1}, 705--719\relax
\mciteBstWouldAddEndPuncttrue
\mciteSetBstMidEndSepPunct{\mcitedefaultmidpunct}
{\mcitedefaultendpunct}{\mcitedefaultseppunct}\relax
\EndOfBibitem
\bibitem[Sullivan and Nienhaus(2025)Sullivan, and Nienhaus]{Sullivan2025AcrossAnnihilation}
Sullivan,~C.~M.; Nienhaus,~L. {Across the Interface: Understanding the Mechanism of Perovskite-Sensitized Triplet–Triplet Annihilation}. \emph{ACS Applied Energy Materials} \textbf{2025}, \relax
\mciteBstWouldAddEndPunctfalse
\mciteSetBstMidEndSepPunct{\mcitedefaultmidpunct}
{}{\mcitedefaultseppunct}\relax
\EndOfBibitem
\bibitem[Huang \latin{et~al.}(2015)Huang, Li, Mahboub, Hanson, Nichols, Le, Tang, and Bardeen]{Huang2015HybridNear-Infrared}
Huang,~Z.; Li,~X.; Mahboub,~M.; Hanson,~K.~M.; Nichols,~V.~M.; Le,~H.; Tang,~M.~L.; Bardeen,~C.~J. {Hybrid Molecule-Nanocrystal Photon Upconversion Across the Visible and Near-Infrared}. \emph{Nano Letters} \textbf{2015}, \emph{15}, 5552--5557\relax
\mciteBstWouldAddEndPuncttrue
\mciteSetBstMidEndSepPunct{\mcitedefaultmidpunct}
{\mcitedefaultendpunct}{\mcitedefaultseppunct}\relax
\EndOfBibitem
\bibitem[Wu \latin{et~al.}(2016)Wu, Congreve, Wilson, Jean, Geva, Welborn, Van~Voorhis, Bulovi{\'{c}}, Bawendi, and Baldo]{Wu2016Solid-stateNanocrystals}
Wu,~M.; Congreve,~D.~N.; Wilson,~M. W.~B.; Jean,~J.; Geva,~N.; Welborn,~M.; Van~Voorhis,~T.; Bulovi{\'{c}},~V.; Bawendi,~M.~G.; Baldo,~M.~A. {Solid-state infrared-to-visible upconversion sensitized by colloidal nanocrystals}. \emph{Nature Photonics} \textbf{2016}, \emph{10}, 31--34\relax
\mciteBstWouldAddEndPuncttrue
\mciteSetBstMidEndSepPunct{\mcitedefaultmidpunct}
{\mcitedefaultendpunct}{\mcitedefaultseppunct}\relax
\EndOfBibitem
\bibitem[Mase \latin{et~al.}(2017)Mase, Okumura, Yanai, and Kimizuka]{Mase2017TripletUpconversion}
Mase,~K.; Okumura,~K.; Yanai,~N.; Kimizuka,~N. {Triplet sensitization by perovskite nanocrystals for photon upconversion}. \emph{Chemical Communications} \textbf{2017}, \emph{53}, 8261--8264\relax
\mciteBstWouldAddEndPuncttrue
\mciteSetBstMidEndSepPunct{\mcitedefaultmidpunct}
{\mcitedefaultendpunct}{\mcitedefaultseppunct}\relax
\EndOfBibitem
\bibitem[Chakkamalayath \latin{et~al.}(2024)Chakkamalayath, Martin, and Kamat]{Chakkamalayath2024EnergyTransfer}
Chakkamalayath,~J.; Martin,~L.~E.; Kamat,~P.~V. {Energy Cascade in Halide Perovskite-Multiple Chromophore Films: Direct versus Mediated Transfer}. \emph{ACS Photonics} \textbf{2024}, \emph{11}, 1821--1831\relax
\mciteBstWouldAddEndPuncttrue
\mciteSetBstMidEndSepPunct{\mcitedefaultmidpunct}
{\mcitedefaultendpunct}{\mcitedefaultseppunct}\relax
\EndOfBibitem
\bibitem[Chakkamalayath and Kamat(2024)Chakkamalayath, and Kamat]{Chakkamalayath2024DemystifyingSystem}
Chakkamalayath,~J.; Kamat,~P.~V. {Demystifying Triplet–Triplet Annihilation Mechanism in the CsPbI$_3$–Rubrene–DBP Upconversion System}. \emph{Journal of the American Chemical Society} \textbf{2024}, \emph{146}, 18095--18103\relax
\mciteBstWouldAddEndPuncttrue
\mciteSetBstMidEndSepPunct{\mcitedefaultmidpunct}
{\mcitedefaultendpunct}{\mcitedefaultseppunct}\relax
\EndOfBibitem
\bibitem[Sullivan and Nienhaus(2022)Sullivan, and Nienhaus]{Sullivan2022RechargingReplacement}
Sullivan,~C.~M.; Nienhaus,~L. {Recharging upconversion: revealing rubrene's replacement}. \emph{Nanoscale} \textbf{2022}, \emph{14}, 17254--17261\relax
\mciteBstWouldAddEndPuncttrue
\mciteSetBstMidEndSepPunct{\mcitedefaultmidpunct}
{\mcitedefaultendpunct}{\mcitedefaultseppunct}\relax
\EndOfBibitem
\bibitem[Sullivan and Nienhaus(2024)Sullivan, and Nienhaus]{Sullivan2024TurningCoupling}
Sullivan,~C.~M.; Nienhaus,~L. {Turning on TTA: Tuning the Energy Landscape by Intermolecular Coupling}. \emph{Chemistry of Materials} \textbf{2024}, \emph{36}, 417--424\relax
\mciteBstWouldAddEndPuncttrue
\mciteSetBstMidEndSepPunct{\mcitedefaultmidpunct}
{\mcitedefaultendpunct}{\mcitedefaultseppunct}\relax
\EndOfBibitem
\bibitem[Sullivan \latin{et~al.}(2024)Sullivan, Szucs, Cantrell, Shulenberger, Siegrist, and Nienhaus]{Sullivan2024WhichAnnihilation}
Sullivan,~C.~M.; Szucs,~A.~M.; Cantrell,~A.~P.; Shulenberger,~K.~E.; Siegrist,~T.; Nienhaus,~L. {Which Flavor of 9,10-Bis(phenylethynyl)Anthracene is Best for Perovskite-Sensitized Triplet–Triplet Annihilation?} \emph{Advanced Energy Materials} \textbf{2024}, 2404130\relax
\mciteBstWouldAddEndPuncttrue
\mciteSetBstMidEndSepPunct{\mcitedefaultmidpunct}
{\mcitedefaultendpunct}{\mcitedefaultseppunct}\relax
\EndOfBibitem
\bibitem[Wieghold \latin{et~al.}(2019)Wieghold, Bieber, VanOrman, and Nienhaus]{Wieghold2019InfluenceUpconversion}
Wieghold,~S.; Bieber,~A.~S.; VanOrman,~Z.~A.; Nienhaus,~L. {Influence of Triplet Diffusion on Lead Halide Perovskite-Sensitized Solid-State Upconversion}. \emph{The Journal of Physical Chemistry Letters} \textbf{2019}, \emph{10}, 3806--3811\relax
\mciteBstWouldAddEndPuncttrue
\mciteSetBstMidEndSepPunct{\mcitedefaultmidpunct}
{\mcitedefaultendpunct}{\mcitedefaultseppunct}\relax
\EndOfBibitem
\bibitem[Wieghold and Nienhaus(2020)Wieghold, and Nienhaus]{Wieghold2020PrechargingDevices}
Wieghold,~S.; Nienhaus,~L. {Precharging Photon Upconversion: Interfacial Interactions in Solution-Processed Perovskite Upconversion Devices}. \emph{Journal of Physical Chemistry Letters} \textbf{2020}, \emph{11}, 601--607\relax
\mciteBstWouldAddEndPuncttrue
\mciteSetBstMidEndSepPunct{\mcitedefaultmidpunct}
{\mcitedefaultendpunct}{\mcitedefaultseppunct}\relax
\EndOfBibitem
\bibitem[Wieghold \latin{et~al.}(2020)Wieghold, Bieber, VanOrman, Rodriguez, and Nienhaus]{Wieghold2020IsRubrene}
Wieghold,~S.; Bieber,~A.~S.; VanOrman,~Z.~A.; Rodriguez,~A.; Nienhaus,~L. {Is Disorder Beneficial in Perovskite-Sensitized Solid-State Upconversion? The Role of DBP Doping in Rubrene}. \emph{The Journal of Physical Chemistry C} \textbf{2020}, \emph{124}, 18132--18140\relax
\mciteBstWouldAddEndPuncttrue
\mciteSetBstMidEndSepPunct{\mcitedefaultmidpunct}
{\mcitedefaultendpunct}{\mcitedefaultseppunct}\relax
\EndOfBibitem
\bibitem[Wieghold \latin{et~al.}(2020)Wieghold, Bieber, Lackner, Nienhaus, Nienhaus, and Nienhaus]{Wieghold2020One-StepDevices}
Wieghold,~S.; Bieber,~A.~S.; Lackner,~J.; Nienhaus,~K.; Nienhaus,~G.~U.; Nienhaus,~L. {One-Step Fabrication of Perovskite-Based Upconversion Devices}. \emph{ChemPhotoChem} \textbf{2020}, \emph{4}, 704--712\relax
\mciteBstWouldAddEndPuncttrue
\mciteSetBstMidEndSepPunct{\mcitedefaultmidpunct}
{\mcitedefaultendpunct}{\mcitedefaultseppunct}\relax
\EndOfBibitem
\bibitem[Bieber \latin{et~al.}(2020)Bieber, VanOrman, Wieghold, and Nienhaus]{Bieber2020Perovskite-sensitizedTemperature}
Bieber,~A.~S.; VanOrman,~Z.~A.; Wieghold,~S.; Nienhaus,~L. {Perovskite-sensitized upconversion bingo: Stoichiometry, composition, solvent, or temperature?} \emph{The Journal of Chemical Physics} \textbf{2020}, \emph{153}, 084703\relax
\mciteBstWouldAddEndPuncttrue
\mciteSetBstMidEndSepPunct{\mcitedefaultmidpunct}
{\mcitedefaultendpunct}{\mcitedefaultseppunct}\relax
\EndOfBibitem
\bibitem[Prashanthan \latin{et~al.}(2020)Prashanthan, Naydenov, Lips, Unger, and MacQueen]{Prashanthan2020InterdependenceAnnihilators}
Prashanthan,~K.; Naydenov,~B.; Lips,~K.; Unger,~E.; MacQueen,~R.~W. {Interdependence of photon upconversion performance and antisolvent processing in thin-film halide perovskite-sensitized triplet–triplet annihilators}. \emph{The Journal of Chemical Physics} \textbf{2020}, \emph{153}, 164711\relax
\mciteBstWouldAddEndPuncttrue
\mciteSetBstMidEndSepPunct{\mcitedefaultmidpunct}
{\mcitedefaultendpunct}{\mcitedefaultseppunct}\relax
\EndOfBibitem
\bibitem[VanOrman \latin{et~al.}(2021)VanOrman, Lackner, Wieghold, Nienhaus, Nienhaus, and Nienhaus]{VanOrman2021EfficiencyMatters}
VanOrman,~Z.~A.; Lackner,~J.; Wieghold,~S.; Nienhaus,~K.; Nienhaus,~G.~U.; Nienhaus,~L. {Efficiency of bulk perovskite-sensitized upconversion: Illuminating matters}. \emph{Applied Physics Letters} \textbf{2021}, \emph{118}, 203903\relax
\mciteBstWouldAddEndPuncttrue
\mciteSetBstMidEndSepPunct{\mcitedefaultmidpunct}
{\mcitedefaultendpunct}{\mcitedefaultseppunct}\relax
\EndOfBibitem
\bibitem[Bieber \latin{et~al.}(2021)Bieber, Vanorman, Drozdick, Weiss, Wieghold, and Nienhaus]{Bieber2021MixedPhonons}
Bieber,~A.~S.; Vanorman,~Z.~A.; Drozdick,~H.~K.; Weiss,~R.; Wieghold,~S.; Nienhaus,~L. {Mixed halide bulk perovskite triplet sensitizers: Interplay between band alignment, mid-gap traps, and phonons}. \emph{Journal of Chemical Physics} \textbf{2021}, \emph{155}, 234706\relax
\mciteBstWouldAddEndPuncttrue
\mciteSetBstMidEndSepPunct{\mcitedefaultmidpunct}
{\mcitedefaultendpunct}{\mcitedefaultseppunct}\relax
\EndOfBibitem
\bibitem[Wang \latin{et~al.}(2021)Wang, Yoo, Lin, Perkinson, Lu, Baldo, and Bawendi]{Wang2021InterfacialUpconversion}
Wang,~L.; Yoo,~J.~J.; Lin,~T.; Perkinson,~C.~F.; Lu,~Y.; Baldo,~M.~A.; Bawendi,~M.~G. {Interfacial Trap‐Assisted Triplet Generation in Lead Halide Perovskite Sensitized Solid‐State Upconversion}. \emph{Advanced Materials} \textbf{2021}, \emph{33}, 2100854\relax
\mciteBstWouldAddEndPuncttrue
\mciteSetBstMidEndSepPunct{\mcitedefaultmidpunct}
{\mcitedefaultendpunct}{\mcitedefaultseppunct}\relax
\EndOfBibitem
\bibitem[Conti \latin{et~al.}(2022)Conti, Bieber, VanOrman, Moller, Wieghold, Schaller, Strouse, and Nienhaus]{Conti2022UltrafastInterface}
Conti,~C.~R.; Bieber,~A.~S.; VanOrman,~Z.~A.; Moller,~G.; Wieghold,~S.; Schaller,~R.~D.; Strouse,~G.~F.; Nienhaus,~L. {Ultrafast Triplet Generation at the Lead Halide Perovskite/Rubrene Interface}. \emph{ACS Energy Letters} \textbf{2022}, \emph{2022}, 617--623\relax
\mciteBstWouldAddEndPuncttrue
\mciteSetBstMidEndSepPunct{\mcitedefaultmidpunct}
{\mcitedefaultendpunct}{\mcitedefaultseppunct}\relax
\EndOfBibitem
\bibitem[Prashanthan \latin{et~al.}(2023)Prashanthan, Levine, Musiienko, Gutierrez-Partida, Hempel, Lips, Unold, Stolterfoht, Dittrich, and MacQueen]{Prashanthan2023InternalUpconverters}
Prashanthan,~K.; Levine,~I.; Musiienko,~A.; Gutierrez-Partida,~E.; Hempel,~H.; Lips,~K.; Unold,~T.; Stolterfoht,~M.; Dittrich,~T.; MacQueen,~R.~W. {Internal electric fields control triplet formation in halide perovskite-sensitized photon upconverters}. \emph{iScience} \textbf{2023}, \emph{26}, 106365\relax
\mciteBstWouldAddEndPuncttrue
\mciteSetBstMidEndSepPunct{\mcitedefaultmidpunct}
{\mcitedefaultendpunct}{\mcitedefaultseppunct}\relax
\EndOfBibitem
\bibitem[Sullivan \latin{et~al.}(2023)Sullivan, Bieber, Drozdick, Moller, Kuszynski, VanOrman, Wieghold, Strouse, and Nienhaus]{Sullivan2023SurfaceUpconversion}
Sullivan,~C.~M.; Bieber,~A.~S.; Drozdick,~H.~K.; Moller,~G.; Kuszynski,~J.~E.; VanOrman,~Z.~A.; Wieghold,~S.; Strouse,~G.~F.; Nienhaus,~L. {Surface Doping Boosts Triplet Generation Yield in Perovskite‐Sensitized Upconversion}. \emph{Advanced Optical Materials} \textbf{2023}, \emph{11}, 2201921\relax
\mciteBstWouldAddEndPuncttrue
\mciteSetBstMidEndSepPunct{\mcitedefaultmidpunct}
{\mcitedefaultendpunct}{\mcitedefaultseppunct}\relax
\EndOfBibitem
\bibitem[Okumoto \latin{et~al.}(2006)Okumoto, Kanno, Hamada, Takahashi, and Shibata]{Okumoto2006HighLayer}
Okumoto,~K.; Kanno,~H.; Hamada,~Y.; Takahashi,~H.; Shibata,~K. {High efficiency red organic light-emitting devices using tetraphenyldibenzoperiflanthene-doped rubrene as an emitting layer}. \emph{Applied Physics Letters} \textbf{2006}, \emph{89}, 013502\relax
\mciteBstWouldAddEndPuncttrue
\mciteSetBstMidEndSepPunct{\mcitedefaultmidpunct}
{\mcitedefaultendpunct}{\mcitedefaultseppunct}\relax
\EndOfBibitem
\bibitem[Bossanyi \latin{et~al.}(2022)Bossanyi, Sasaki, Wang, Chekulaev, Kimizuka, Yanai, and Clark]{Bossanyi2022InFission}
Bossanyi,~D.~G.; Sasaki,~Y.; Wang,~S.; Chekulaev,~D.; Kimizuka,~N.; Yanai,~N.; Clark,~J. {In optimized rubrene-based nanoparticle blends for photon upconversion, singlet energy collection outcompetes triplet-pair separation, not singlet fission}. \emph{Journal of Materials Chemistry C} \textbf{2022}, \emph{10}, 4684--4696\relax
\mciteBstWouldAddEndPuncttrue
\mciteSetBstMidEndSepPunct{\mcitedefaultmidpunct}
{\mcitedefaultendpunct}{\mcitedefaultseppunct}\relax
\EndOfBibitem
\bibitem[De~Wolf \latin{et~al.}(2014)De~Wolf, Holovsky, Moon, L{\"{o}}per, Niesen, Ledinsky, Haug, Yum, and Ballif]{DeWolf2014OrganometallicPerformance}
De~Wolf,~S.; Holovsky,~J.; Moon,~S.~J.; L{\"{o}}per,~P.; Niesen,~B.; Ledinsky,~M.; Haug,~F.~J.; Yum,~J.~H.; Ballif,~C. {Organometallic halide perovskites: Sharp optical absorption edge and its relation to photovoltaic performance}. \emph{Journal of Physical Chemistry Letters} \textbf{2014}, \emph{5}, 1035--1039\relax
\mciteBstWouldAddEndPuncttrue
\mciteSetBstMidEndSepPunct{\mcitedefaultmidpunct}
{\mcitedefaultendpunct}{\mcitedefaultseppunct}\relax
\EndOfBibitem
\bibitem[Bailey \latin{et~al.}(2019)Bailey, Piana, and Lagoudakis]{Bailey2019High-EnergySpectrophotometry}
Bailey,~C.~G.; Piana,~G.~M.; Lagoudakis,~P.~G. {High-Energy Optical Transitions and Optical Constants of CH$_3$NH$_3$PbI$_3$ Measured by Spectroscopic Ellipsometry and Spectrophotometry}. \emph{The Journal of Physical Chemistry C} \textbf{2019}, \emph{123}, 28795--28801\relax
\mciteBstWouldAddEndPuncttrue
\mciteSetBstMidEndSepPunct{\mcitedefaultmidpunct}
{\mcitedefaultendpunct}{\mcitedefaultseppunct}\relax
\EndOfBibitem
\bibitem[Ponseca \latin{et~al.}(2014)Ponseca, Savenije, Abdellah, Zheng, Yartsev, Pascher, Harlang, Chabera, Pullerits, Stepanov, Wolf, and Sundstr{\"{o}}m]{Ponseca2014OrganometalRecombination}
Ponseca,~C.~S.; Savenije,~T.~J.; Abdellah,~M.; Zheng,~K.; Yartsev,~A.; Pascher,~T.; Harlang,~T.; Chabera,~P.; Pullerits,~T.; Stepanov,~A.; Wolf,~J.~P.; Sundstr{\"{o}}m,~V. {Organometal halide perovskite solar cell materials rationalized: Ultrafast charge generation, high and microsecond-long balanced mobilities, and slow recombination}. \emph{Journal of the American Chemical Society} \textbf{2014}, \emph{136}, 5189--5192\relax
\mciteBstWouldAddEndPuncttrue
\mciteSetBstMidEndSepPunct{\mcitedefaultmidpunct}
{\mcitedefaultendpunct}{\mcitedefaultseppunct}\relax
\EndOfBibitem
\bibitem[Galkowski \latin{et~al.}(2016)Galkowski, Mitioglu, Miyata, Plochocka, Portugall, Eperon, Wang, Stergiopoulos, Stranks, Snaith, and Nicholas]{Galkowski2016DeterminationSemiconductors}
Galkowski,~K.; Mitioglu,~A.; Miyata,~A.; Plochocka,~P.; Portugall,~O.; Eperon,~G.~E.; Wang,~J. T.~W.; Stergiopoulos,~T.; Stranks,~S.~D.; Snaith,~H.~J.; Nicholas,~R.~J. {Determination of the exciton binding energy and effective masses for methylammonium and formamidinium lead tri-halide perovskite semiconductors}. \emph{Energy and Environmental Science} \textbf{2016}, \emph{9}, 962--970\relax
\mciteBstWouldAddEndPuncttrue
\mciteSetBstMidEndSepPunct{\mcitedefaultmidpunct}
{\mcitedefaultendpunct}{\mcitedefaultseppunct}\relax
\EndOfBibitem
\bibitem[Alves \latin{et~al.}(2022)Alves, Feng, Nienhaus, and Schmidt]{Alves2022ChallengesUpconversion}
Alves,~J.; Feng,~J.; Nienhaus,~L.; Schmidt,~T.~W. {Challenges, progress and prospects in solid state triplet fusion upconversion}. \emph{Journal of Materials Chemistry C} \textbf{2022}, \emph{10}, 7783--7798\relax
\mciteBstWouldAddEndPuncttrue
\mciteSetBstMidEndSepPunct{\mcitedefaultmidpunct}
{\mcitedefaultendpunct}{\mcitedefaultseppunct}\relax
\EndOfBibitem
\bibitem[Lin \latin{et~al.}(2020)Lin, Perkinson, Baldo, Lin, Baldo, and Perkinson]{Lin2020StrategiesUpconversion}
Lin,~T.-A.; Perkinson,~C.~F.; Baldo,~M.~A.; Lin,~T.-a.; Baldo,~M.~A.; Perkinson,~C.~F. {Strategies for High-Performance Solid-State Triplet–Triplet-Annihilation-Based Photon Upconversion}. \emph{Advanced Materials} \textbf{2020}, \emph{32}, 1908175\relax
\mciteBstWouldAddEndPuncttrue
\mciteSetBstMidEndSepPunct{\mcitedefaultmidpunct}
{\mcitedefaultendpunct}{\mcitedefaultseppunct}\relax
\EndOfBibitem
\bibitem[Narayanan \latin{et~al.}(2024)Narayanan, Hu, Gallegos, Pucurimay, Zhou, Belliveau, Ahmed, Fern{\'{a}}ndez, Michaels, Murrietta, Mutatu, Feng, Hamid, Yap, Schloemer, Jaramillo, Kats, and Congreve]{Narayanan2024OvercomingUpconversion}
Narayanan,~P.; Hu,~M.; Gallegos,~A.~O.; Pucurimay,~L.; Zhou,~Q.; Belliveau,~E.; Ahmed,~G.~H.; Fern{\'{a}}ndez,~S.; Michaels,~W.; Murrietta,~N.; Mutatu,~V.~E.; Feng,~D.; Hamid,~R.; Yap,~K. M.~K.; Schloemer,~T.~H.; Jaramillo,~T.~F.; Kats,~M.~A.; Congreve,~D.~N. {Overcoming the Absorption Bottleneck for Solid-State Infrared-to-Visible Upconversion}. \emph{ChemRxiv} \textbf{2024}, DOI: 10.26434/chemrxiv--2024--h0k05\relax
\mciteBstWouldAddEndPuncttrue
\mciteSetBstMidEndSepPunct{\mcitedefaultmidpunct}
{\mcitedefaultendpunct}{\mcitedefaultseppunct}\relax
\EndOfBibitem
\bibitem[Wu \latin{et~al.}(2022)Wu, Liang, Ge, Sun, Zhang, and Xing]{Wu2022SurfaceCells}
Wu,~G.; Liang,~R.; Ge,~M.; Sun,~G.; Zhang,~Y.; Xing,~G. {Surface Passivation Using 2D Perovskites toward Efficient and Stable Perovskite Solar Cells}. \emph{Advanced Materials} \textbf{2022}, \emph{34}, 2105635\relax
\mciteBstWouldAddEndPuncttrue
\mciteSetBstMidEndSepPunct{\mcitedefaultmidpunct}
{\mcitedefaultendpunct}{\mcitedefaultseppunct}\relax
\EndOfBibitem
\bibitem[Saliba \latin{et~al.}(2016)Saliba, Matsui, Seo, Domanski, Correa-Baena, Nazeeruddin, Zakeeruddin, Tress, Abate, Hagfeldt, and Gr{\"{a}}tzel]{Saliba2016Cesium-containingEfficiency}
Saliba,~M.; Matsui,~T.; Seo,~J.~Y.; Domanski,~K.; Correa-Baena,~J.~P.; Nazeeruddin,~M.~K.; Zakeeruddin,~S.~M.; Tress,~W.; Abate,~A.; Hagfeldt,~A.; Gr{\"{a}}tzel,~M. {Cesium-containing triple cation perovskite solar cells: improved stability, reproducibility and high efficiency}. \emph{Energy {\&} Environmental Science} \textbf{2016}, \emph{9}, 1989--1997\relax
\mciteBstWouldAddEndPuncttrue
\mciteSetBstMidEndSepPunct{\mcitedefaultmidpunct}
{\mcitedefaultendpunct}{\mcitedefaultseppunct}\relax
\EndOfBibitem
\bibitem[Chen \latin{et~al.}(2018)Chen, Bai, Wang, Lyu, Yun, and Wang]{Chen2018}
Chen,~P.; Bai,~Y.; Wang,~S.; Lyu,~M.; Yun,~J.~H.; Wang,~L. {In Situ Growth of 2D Perovskite Capping Layer for Stable and Efficient Perovskite Solar Cells}. \emph{Advanced Functional Materials} \textbf{2018}, \emph{28}, 1706923\relax
\mciteBstWouldAddEndPuncttrue
\mciteSetBstMidEndSepPunct{\mcitedefaultmidpunct}
{\mcitedefaultendpunct}{\mcitedefaultseppunct}\relax
\EndOfBibitem
\bibitem[Cho \latin{et~al.}(2018)Cho, Grancini, Lee, Oveisi, Ryu, Almora, Tschumi, Schouwink, Seo, Heo, Park, Jang, Paek, Garcia-Belmonte, and Nazeeruddin]{Cho2018SelectivePhotovoltaics}
Cho,~K.~T.; Grancini,~G.; Lee,~Y.; Oveisi,~E.; Ryu,~J.; Almora,~O.; Tschumi,~M.; Schouwink,~P.~A.; Seo,~G.; Heo,~S.; Park,~J.; Jang,~J.; Paek,~S.; Garcia-Belmonte,~G.; Nazeeruddin,~M.~K. {Selective growth of layered perovskites for stable and efficient photovoltaics}. \emph{Energy {\&} Environmental Science} \textbf{2018}, \emph{11}, 952--959\relax
\mciteBstWouldAddEndPuncttrue
\mciteSetBstMidEndSepPunct{\mcitedefaultmidpunct}
{\mcitedefaultendpunct}{\mcitedefaultseppunct}\relax
\EndOfBibitem
\bibitem[Jang \latin{et~al.}(2021)Jang, Lee, Yeom, Jeong, Choi, Choi, and Noh]{Jang2021IntactGrowth}
Jang,~Y.-W.; Lee,~S.; Yeom,~K.~M.; Jeong,~K.; Choi,~K.; Choi,~M.; Noh,~J.~H. {Intact 2D/3D halide junction perovskite solar cells via solid-phase in-plane growth}. \emph{Nature Energy} \textbf{2021}, \emph{6}, 63--71\relax
\mciteBstWouldAddEndPuncttrue
\mciteSetBstMidEndSepPunct{\mcitedefaultmidpunct}
{\mcitedefaultendpunct}{\mcitedefaultseppunct}\relax
\EndOfBibitem
\bibitem[Sutanto \latin{et~al.}(2021)Sutanto, Caprioglio, Drigo, Hofstetter, Garcia-Benito, Queloz, Neher, Nazeeruddin, Stolterfoht, Vaynzof, and Grancini]{Sutanto20212D/3DCells}
Sutanto,~A.~A.; Caprioglio,~P.; Drigo,~N.; Hofstetter,~Y.~J.; Garcia-Benito,~I.; Queloz,~V.~I.; Neher,~D.; Nazeeruddin,~M.~K.; Stolterfoht,~M.; Vaynzof,~Y.; Grancini,~G. {2D/3D perovskite engineering eliminates interfacial recombination losses in hybrid perovskite solar cells}. \emph{Chem} \textbf{2021}, \emph{7}, 1903--1916\relax
\mciteBstWouldAddEndPuncttrue
\mciteSetBstMidEndSepPunct{\mcitedefaultmidpunct}
{\mcitedefaultendpunct}{\mcitedefaultseppunct}\relax
\EndOfBibitem
\bibitem[Teale \latin{et~al.}(2024)Teale, Degani, Chen, Sargent, and Grancini]{Teale2024MolecularCells}
Teale,~S.; Degani,~M.; Chen,~B.; Sargent,~E.~H.; Grancini,~G. {Molecular cation and low-dimensional perovskite surface passivation in perovskite solar cells}. \emph{Nature Energy} \textbf{2024}, \emph{9}, 779--792\relax
\mciteBstWouldAddEndPuncttrue
\mciteSetBstMidEndSepPunct{\mcitedefaultmidpunct}
{\mcitedefaultendpunct}{\mcitedefaultseppunct}\relax
\EndOfBibitem
\bibitem[Fu \latin{et~al.}(2018)Fu, Zheng, Wang, Hautzinger, Pan, Dang, Wright, Pan, and Jin]{Fu2018MulticolorTransfer}
Fu,~Y.; Zheng,~W.; Wang,~X.; Hautzinger,~M.~P.; Pan,~D.; Dang,~L.; Wright,~J.~C.; Pan,~A.; Jin,~S. {Multicolor Heterostructures of Two-Dimensional Layered Halide Perovskites that Show Interlayer Energy Transfer}. \emph{Journal of the American Chemical Society} \textbf{2018}, \emph{140}, 15675--15683\relax
\mciteBstWouldAddEndPuncttrue
\mciteSetBstMidEndSepPunct{\mcitedefaultmidpunct}
{\mcitedefaultendpunct}{\mcitedefaultseppunct}\relax
\EndOfBibitem
\bibitem[Wen \latin{et~al.}(2023)Wen, Zhao, Wu, Liu, Zheng, Lin, Wan, Li, Luo, Tian, Li, and Tan]{Wen2023HeterojunctionCells}
Wen,~J.; Zhao,~Y.; Wu,~P.; Liu,~Y.; Zheng,~X.; Lin,~R.; Wan,~S.; Li,~K.; Luo,~H.; Tian,~Y.; Li,~L.; Tan,~H. {Heterojunction formed via 3D-to-2D perovskite conversion for photostable wide-bandgap perovskite solar cells}. \emph{Nature Communications} \textbf{2023}, \emph{14}, 7118\relax
\mciteBstWouldAddEndPuncttrue
\mciteSetBstMidEndSepPunct{\mcitedefaultmidpunct}
{\mcitedefaultendpunct}{\mcitedefaultseppunct}\relax
\EndOfBibitem
\bibitem[Merrifield(1968)]{Merrifield1968TheoryExcitons}
Merrifield,~R.~E. {Theory of Magnetic Field Effects on the Mutual Annihilation of Triplet Excitons}. \emph{The Journal of Chemical Physics} \textbf{1968}, \emph{48}, 4318--4319\relax
\mciteBstWouldAddEndPuncttrue
\mciteSetBstMidEndSepPunct{\mcitedefaultmidpunct}
{\mcitedefaultendpunct}{\mcitedefaultseppunct}\relax
\EndOfBibitem
\bibitem[Johnson and Merrifield(1970)Johnson, and Merrifield]{Johnson1970EffectsCrystals}
Johnson,~R.~C.; Merrifield,~R.~E. {Effects of Magnetic Fields on the Mutual Annihilation of Triplet Excitons in Anthracene Crystals}. \emph{Physical Review B} \textbf{1970}, \emph{1}, 896--902\relax
\mciteBstWouldAddEndPuncttrue
\mciteSetBstMidEndSepPunct{\mcitedefaultmidpunct}
{\mcitedefaultendpunct}{\mcitedefaultseppunct}\relax
\EndOfBibitem
\bibitem[Merrifield(1971)]{Merrifield1971MagneticInteractions}
Merrifield,~R.~E. {Magnetic effects on triplet exciton interactions}. \emph{Pure and Applied Chemistry} \textbf{1971}, \emph{27}, 481--498\relax
\mciteBstWouldAddEndPuncttrue
\mciteSetBstMidEndSepPunct{\mcitedefaultmidpunct}
{\mcitedefaultendpunct}{\mcitedefaultseppunct}\relax
\EndOfBibitem
\bibitem[Ern and Merrifield(1968)Ern, and Merrifield]{Ern1968MagneticCrystals}
Ern,~V.; Merrifield,~R.~E. {Magnetic Field Effect on Triplet Exciton Quenching in Organic Crystals}. \emph{Physical Review Letters} \textbf{1968}, \emph{21}, 609--611\relax
\mciteBstWouldAddEndPuncttrue
\mciteSetBstMidEndSepPunct{\mcitedefaultmidpunct}
{\mcitedefaultendpunct}{\mcitedefaultseppunct}\relax
\EndOfBibitem
\bibitem[Swenburg \latin{et~al.}(1973)Swenburg, Geacintov, and Birks]{Swenburg1973OrganicPhotophysics}
Swenburg,~C.~E.; Geacintov,~B.~E.; Birks,~J.~B. In \emph{{Organic molecular photophysics}}; Birks,~J.~B., Ed.; 1973; Vol.~1; pp 521--523\relax
\mciteBstWouldAddEndPuncttrue
\mciteSetBstMidEndSepPunct{\mcitedefaultmidpunct}
{\mcitedefaultendpunct}{\mcitedefaultseppunct}\relax
\EndOfBibitem
\bibitem[Shao \latin{et~al.}(2013)Shao, Yan, Li, Ilia, and Hu]{Shao2013TripletchargeSemiconductorsb}
Shao,~M.; Yan,~L.; Li,~M.; Ilia,~I.; Hu,~B. {Triplet–charge annihilation versus triplet–triplet annihilation in organic semiconductors}. \emph{J. Mater. Chem. C} \textbf{2013}, \emph{1}, 1330--1336\relax
\mciteBstWouldAddEndPuncttrue
\mciteSetBstMidEndSepPunct{\mcitedefaultmidpunct}
{\mcitedefaultendpunct}{\mcitedefaultseppunct}\relax
\EndOfBibitem
\bibitem[Thompson \latin{et~al.}(2014)Thompson, Hontz, Congreve, Bahlke, Reineke, Van~Voorhis, and Baldo]{Thompson2014NanostructuredAnnihilation}
Thompson,~N.~J.; Hontz,~E.; Congreve,~D.~N.; Bahlke,~M.~E.; Reineke,~S.; Van~Voorhis,~T.; Baldo,~M.~A. {Nanostructured Singlet Fission Photovoltaics Subject to Triplet‐Charge Annihilation}. \emph{Advanced Materials} \textbf{2014}, \emph{26}, 1366--1371\relax
\mciteBstWouldAddEndPuncttrue
\mciteSetBstMidEndSepPunct{\mcitedefaultmidpunct}
{\mcitedefaultendpunct}{\mcitedefaultseppunct}\relax
\EndOfBibitem
\bibitem[Ji \latin{et~al.}(2017)Ji, Zheng, Zhao, Song, Zhang, Shen, Yang, Xiong, Gao, Cao, and Qi]{Ji2017InterfacialInterface}
Ji,~G.; Zheng,~G.; Zhao,~B.; Song,~F.; Zhang,~X.; Shen,~K.; Yang,~Y.; Xiong,~Y.; Gao,~X.; Cao,~L.; Qi,~D.-c. {Interfacial electronic structures revealed at the rubrene/CH$_3$NH$_3$PbI$_3$ interface}. \emph{Physical Chemistry Chemical Physics} \textbf{2017}, \emph{19}, 6546--6553\relax
\mciteBstWouldAddEndPuncttrue
\mciteSetBstMidEndSepPunct{\mcitedefaultmidpunct}
{\mcitedefaultendpunct}{\mcitedefaultseppunct}\relax
\EndOfBibitem
\bibitem[Sloane \latin{et~al.}(2025)Sloane, Bailey, Cole, Schmidt, McCamey, and Klymenko]{Sloane2025ElectronicTermination}
Sloane,~N.~P.; Bailey,~C.~G.; Cole,~J.~H.; Schmidt,~T.~W.; McCamey,~D.~R.; Klymenko,~M.~V. {Electronic Structure at the Perovskite/Rubrene Interface: The Effect of Surface Termination}. \emph{The Journal of Physical Chemistry C} \textbf{2025}, \emph{129}, 889--898\relax
\mciteBstWouldAddEndPuncttrue
\mciteSetBstMidEndSepPunct{\mcitedefaultmidpunct}
{\mcitedefaultendpunct}{\mcitedefaultseppunct}\relax
\EndOfBibitem
\bibitem[Tang \latin{et~al.}(2024)Tang, Lin, Yan, Lin, Rao, Pan, and Zhong]{Tang202420.1Perovskites}
Tang,~J.; Lin,~Y.; Yan,~H.; Lin,~J.; Rao,~H.; Pan,~Z.; Zhong,~X. {20.1 {\%} Certified Efficiency of Planar Hole Transport Layer‐Free Carbon‐Based Perovskite Solar Cells by Spacer Cation Chain Length Engineering of 2D Perovskites}. \emph{Angewandte Chemie International Edition} \textbf{2024}, \emph{63}, e202406167\relax
\mciteBstWouldAddEndPuncttrue
\mciteSetBstMidEndSepPunct{\mcitedefaultmidpunct}
{\mcitedefaultendpunct}{\mcitedefaultseppunct}\relax
\EndOfBibitem
\bibitem[Bailey \latin{et~al.}(2024)Bailey, Gillan, Lee, Sloane, Liu, Hao, Soufiani, and McCamey]{Bailey2024InfluencePerovskites}
Bailey,~C.~G.; Gillan,~L.~V.; Lee,~M.; Sloane,~N.; Liu,~X.; Hao,~X.; Soufiani,~A.~M.; McCamey,~D.~R. {Influence of Organic Spacer Cation on Dark Excitons in 2D Perovskites}. \emph{Advanced Functional Materials} \textbf{2024}, \emph{34}, 2308095\relax
\mciteBstWouldAddEndPuncttrue
\mciteSetBstMidEndSepPunct{\mcitedefaultmidpunct}
{\mcitedefaultendpunct}{\mcitedefaultseppunct}\relax
\EndOfBibitem
\bibitem[Bailey \latin{et~al.}(2025)Bailey, Mena, Leung, Sloane, Liao, McKenzie, McCamey, and Ho-Baillie]{Bailey2025RevealingMicroscopy}
Bailey,~C.~G.; Mena,~A.; Leung,~T.~L.; Sloane,~N.~P.; Liao,~C.; McKenzie,~D.~R.; McCamey,~D.~R.; Ho-Baillie,~A. {Revealing localised dark-exciton populations in 2D perovskites via magneto-optical microscopy}. \textbf{2025}, \relax
\mciteBstWouldAddEndPunctfalse
\mciteSetBstMidEndSepPunct{\mcitedefaultmidpunct}
{}{\mcitedefaultseppunct}\relax
\EndOfBibitem
\bibitem[Cheng \latin{et~al.}(2009)Cheng, Khoury, Clady, Tayebjee, Ekins-Daukes, Crossley, and Schmidt]{Cheng2009OnUpconversion}
Cheng,~Y.~Y.; Khoury,~T.; Clady,~R.~G.; Tayebjee,~M.~J.; Ekins-Daukes,~N.~J.; Crossley,~M.~J.; Schmidt,~T.~W. {On the efficiency limit of triplet–triplet annihilation for photochemical upconversion}. \emph{Physical Chemistry Chemical Physics} \textbf{2009}, \emph{12}, 66--71\relax
\mciteBstWouldAddEndPuncttrue
\mciteSetBstMidEndSepPunct{\mcitedefaultmidpunct}
{\mcitedefaultendpunct}{\mcitedefaultseppunct}\relax
\EndOfBibitem
\bibitem[Haefele \latin{et~al.}(2012)Haefele, Blumhoff, Khnayzer, and Castellano]{Haefele2012GettingLinear}
Haefele,~A.; Blumhoff,~J.; Khnayzer,~R.~S.; Castellano,~F.~N. {Getting to the (Square) root of the problem: How to make noncoherent pumped upconversion linear}. \emph{Journal of Physical Chemistry Letters} \textbf{2012}, \emph{3}, 299--303\relax
\mciteBstWouldAddEndPuncttrue
\mciteSetBstMidEndSepPunct{\mcitedefaultmidpunct}
{\mcitedefaultendpunct}{\mcitedefaultseppunct}\relax
\EndOfBibitem
\bibitem[deQuilettes \latin{et~al.}(2016)deQuilettes, Zhang, Burlakov, Graham, Leijtens, Osherov, Bulovi{\'{c}}, Snaith, Ginger, and Stranks]{DeQuilettes2016Photo-inducedFilms}
deQuilettes,~D.~W.; Zhang,~W.; Burlakov,~V.~M.; Graham,~D.~J.; Leijtens,~T.; Osherov,~A.; Bulovi{\'{c}},~V.; Snaith,~H.~J.; Ginger,~D.~S.; Stranks,~S.~D. {Photo-induced halide redistribution in organic–inorganic perovskite films}. \emph{Nature Communications} \textbf{2016}, \emph{7}, 11683\relax
\mciteBstWouldAddEndPuncttrue
\mciteSetBstMidEndSepPunct{\mcitedefaultmidpunct}
{\mcitedefaultendpunct}{\mcitedefaultseppunct}\relax
\EndOfBibitem
\bibitem[Wei \latin{et~al.}(2021)Wei, Wang, Huo, Gao, Gan, Zhao, and Li]{Wei2021MechanismsCells}
Wei,~J.; Wang,~Q.; Huo,~J.; Gao,~F.; Gan,~Z.; Zhao,~Q.; Li,~H. {Mechanisms and Suppression of Photoinduced Degradation in Perovskite Solar Cells}. \emph{Advanced Energy Materials} \textbf{2021}, \emph{11}, 2002326\relax
\mciteBstWouldAddEndPuncttrue
\mciteSetBstMidEndSepPunct{\mcitedefaultmidpunct}
{\mcitedefaultendpunct}{\mcitedefaultseppunct}\relax
\EndOfBibitem
\bibitem[Grancini \latin{et~al.}(2017)Grancini, Rold{\'{a}}n-Carmona, Zimmermann, Mosconi, Lee, Martineau, Narbey, Oswald, De~Angelis, Graetzel, and Nazeeruddin]{Grancini2017}
Grancini,~G.; Rold{\'{a}}n-Carmona,~C.; Zimmermann,~I.; Mosconi,~E.; Lee,~X.; Martineau,~D.; Narbey,~S.; Oswald,~F.; De~Angelis,~F.; Graetzel,~M.; Nazeeruddin,~M.~K. {One-Year stable perovskite solar cells by 2D/3D interface engineering}. \emph{Nature Communications} \textbf{2017}, \emph{8}, 1--8\relax
\mciteBstWouldAddEndPuncttrue
\mciteSetBstMidEndSepPunct{\mcitedefaultmidpunct}
{\mcitedefaultendpunct}{\mcitedefaultseppunct}\relax
\EndOfBibitem
\bibitem[Zhao \latin{et~al.}(2022)Zhao, Liu, and Loo]{Zhao2022a}
Zhao,~X.; Liu,~T.; Loo,~Y. {Advancing 2D Perovskites for Efficient and Stable Solar Cells: Challenges and Opportunities}. \emph{Advanced Materials} \textbf{2022}, \emph{34}, 2105849\relax
\mciteBstWouldAddEndPuncttrue
\mciteSetBstMidEndSepPunct{\mcitedefaultmidpunct}
{\mcitedefaultendpunct}{\mcitedefaultseppunct}\relax
\EndOfBibitem
\bibitem[Zhang \latin{et~al.}(2017)Zhang, Wu, Yang, Fu, Zhang, Chen, Liu, Yan, Yang, and Chen]{Zhang2017VerticallyStability}
Zhang,~X.; Wu,~G.; Yang,~S.; Fu,~W.; Zhang,~Z.; Chen,~C.; Liu,~W.; Yan,~J.; Yang,~W.; Chen,~H. {Vertically Oriented 2D Layered Perovskite Solar Cells with Enhanced Efficiency and Good Stability}. \emph{Small} \textbf{2017}, \emph{13}, 1700611\relax
\mciteBstWouldAddEndPuncttrue
\mciteSetBstMidEndSepPunct{\mcitedefaultmidpunct}
{\mcitedefaultendpunct}{\mcitedefaultseppunct}\relax
\EndOfBibitem
\bibitem[Chen \latin{et~al.}(2018)Chen, Shiu, Ma, Alpert, Zhang, Foley, Smilgies, Lee, and Choi]{Chen2018OriginPerformance}
Chen,~A.~Z.; Shiu,~M.; Ma,~J.~H.; Alpert,~M.~R.; Zhang,~D.; Foley,~B.~J.; Smilgies,~D.-M.; Lee,~S.-H.; Choi,~J.~J. {Origin of vertical orientation in two-dimensional metal halide perovskites and its effect on photovoltaic performance}. \emph{Nature Communications} \textbf{2018}, \emph{9}, 1336\relax
\mciteBstWouldAddEndPuncttrue
\mciteSetBstMidEndSepPunct{\mcitedefaultmidpunct}
{\mcitedefaultendpunct}{\mcitedefaultseppunct}\relax
\EndOfBibitem
\end{mcitethebibliography}

\clearpage

\section{Supporting Information}

\setcounter{figure}{0}
\makeatletter 
\renewcommand{\thefigure}{S\@arabic\c@figure}
\makeatother

\setcounter{table}{0}
\makeatletter 
\renewcommand{\thetable}{S\@arabic\c@table}
\makeatother

\begin{figure}[!ht]
    \centering
    \includegraphics[width = \linewidth]{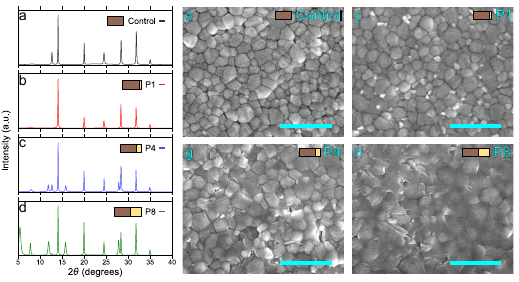}
    \caption{XRD spectra of the neat (a) Control Br17 film, and the PEAI-treated samples (b) P1, (c) P4, and (d) P8. Scanning electron microscope images of (e) Control, and the PEAI-treated samples (f) P1, (g) P4, and (h) P8. The scale bar in figures e-h represents 1 $\mu$m.}
    \label{sfig:xrd_sem}
\end{figure}

\section{Supporting Note 1}

\textbf{Figure \ref{sfig:xrd_sem}} shows the X-ray diffraction (XRD) and Scanning Electron Microscopy (SEM) of all sample types. In \textbf{Figure \ref{sfig:xrd_sem}a-d} the peaks from Br17 can be seen and are present in all XRD spectra for all samples. However, there are two differences of note; the first is the disappearance of the characteristic PbI$_2$ peak at 12.7$^\circ$\textsuperscript{1} for the P1 sample (\textbf{Figure \ref{sfig:xrd_sem}b}) suggesting the reaction of the PbI$_2$ with the PEAI to form the passivation layer. The second is the appearance of additional peaks for the samples P4 and P8 (\textbf{Figure \ref{sfig:xrd_sem}c and d}), which are attributed to the presence of the 2D layer and residual PEAI\textsuperscript{2}, with intensities increasing from P4 to P8 indicating the formation of a thicker spacer layer with increasing PEAI concentration. Similarly, scanning electron microscope (SEM) images of the surface of the films show that with increasing PEAI concentration treatment, the grains begin to fuse as the 2D perovskite layer becomes thicker (\textbf{Figure \ref{sfig:xrd_sem}e-h}). 

\begin{figure}[!ht]
    \centering
    \includegraphics[width= 0.7\linewidth]{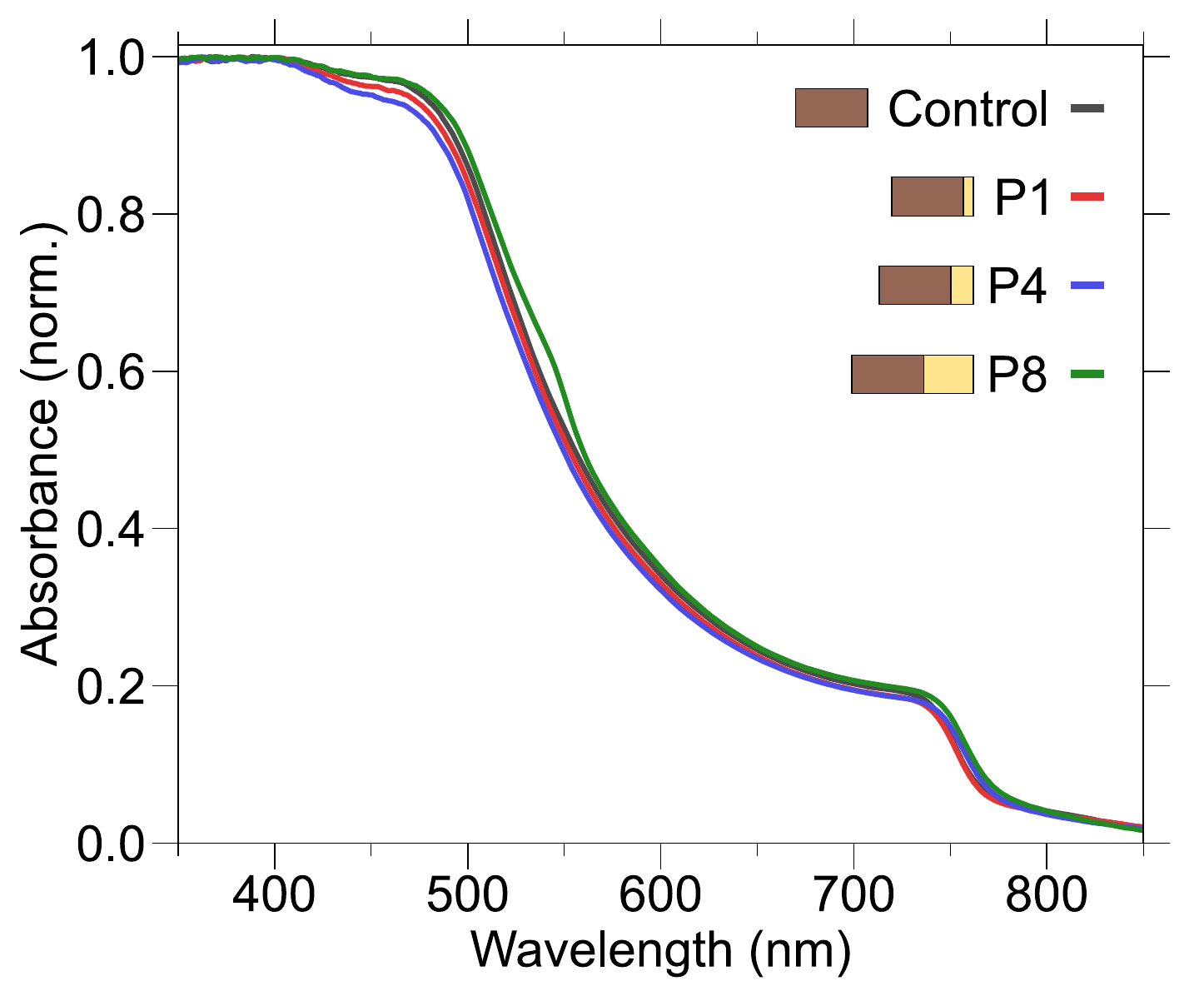}
    \caption{Normalised absorption spectra of the neat Control film (black), alongside the PEAI-treated samples P1 (red), P4 (blue), and P8 (green).}
    \label{sfig:norm_abs}
\end{figure}

\begin{figure}[!ht]
    \centering
    \includegraphics[width=\linewidth]{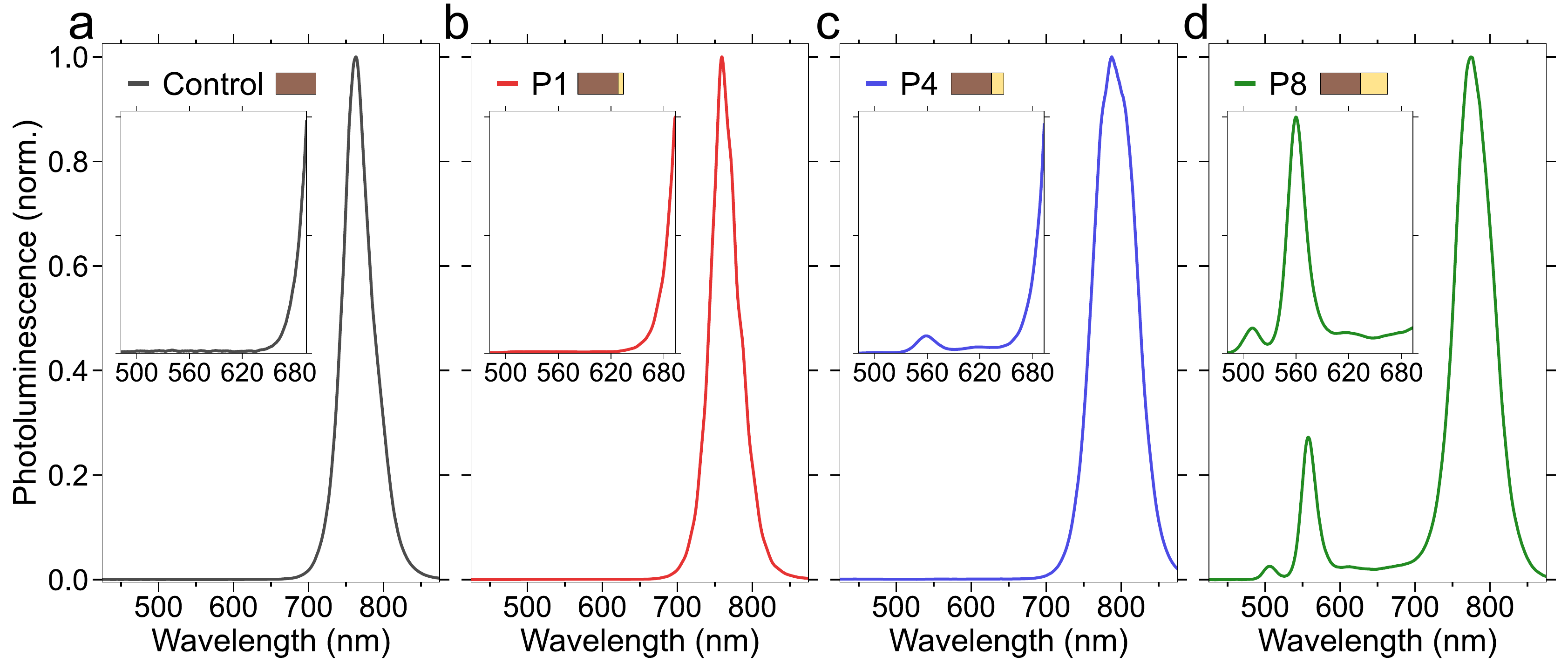}
    \caption{Photoluminescence spectra of the neat (a) Control Br17 film, and the PEAI-treated samples (b) P1, (c) P4, and (d) P8. The inset in each plot is the emission from 475-700\,nm highlighting the emission from the 2D perovskite spacer layer for (c) P4 and (d) P8.}
    \label{sfig:perov-pl}
\end{figure}

\clearpage

\section{Supporting Note 2}
The photoluminescence of all neat perovskite films under 405\,nm excitation is shown in \textbf{Figure \ref{sfig:perov-pl}}, showing Gaussian-like emission of the Br17 centred around 780\,nm. However, for the samples P4 and P8, additional shorter wavelength features appear. The insets in \textbf{Figures \ref{sfig:perov-pl}c} and \textbf{d} show the higher energy emission from the capping layer. From \textbf{Figure \ref{sfig:perov-pl}d}, it can be determined that there are multiple emissive species in the 2D perovskite layer, potentially due to the capping layer consisting of a mixture of compositions due to the multiple cation/halides present in the underlying Br17 layer.\textsuperscript{3}\\

\begin{figure}[!ht]
    \centering
    \includegraphics[width=0.75\linewidth]{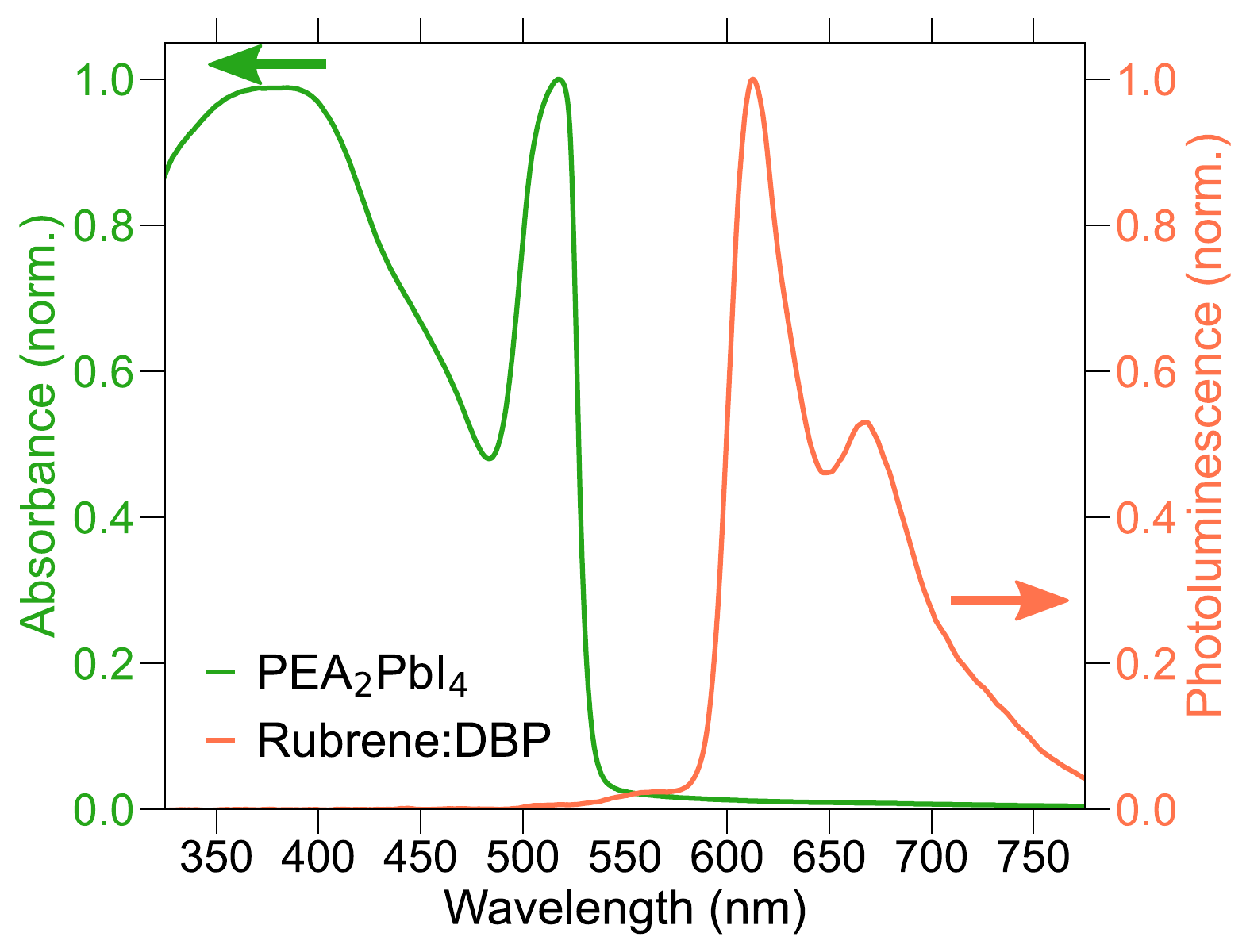}
    \caption{Comparison of the absorbance spectra of PEA$_2$PbI$_4$ (green) and the photoluminescence spectra of Rubrene:DBP (orange) excited at 405\,nm.}
    \label{sfig:pepi_rub}
\end{figure}

\clearpage

\begin{figure}[!ht]
    \centering
    \includegraphics[width=0.75\linewidth]{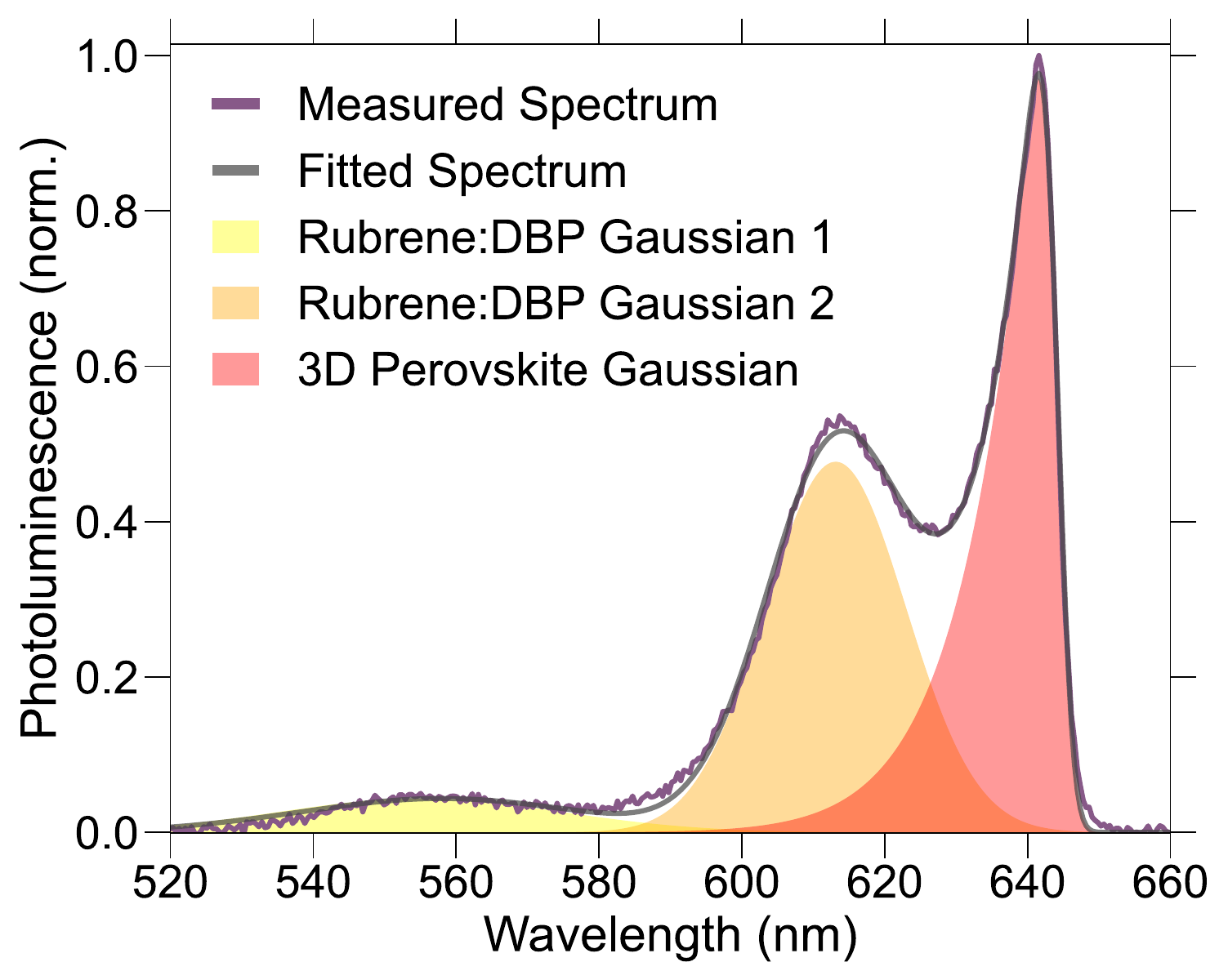}
    \caption{Spectrum decomposition used to determine the intensities of the upconverted emission from the rubrene:DBP film and separate the tail emission of the 3D perovskite cut-off by a 650\,nm short-pass filter. The figure shows an example of the upconverted spectrum from a Control/Rub bilayer sample excited at 670\,nm with the measured spectrum (purple), the fitted spectrum (grey), two gaussians from the rubrene:DBP emission (yellow, orange), and the gaussian of the perovskite PL truncated by a function modelling the transmission through a 650\,nm shortpass filter (red).}
    \label{sfig:spec_decon}
\end{figure}

\clearpage

\section{Supporting Note 3}
Time-Correlated Single Photon Counting (TCSPC) was measured for all films excited at 700\,nm, ensuring charge carriers are photogenerated in the bulk perovskite layer only. Subsequently, their dynamics can be determined from the photoluminescence decay over time. A triexponential function was fit to the decay traces in \textbf{Figure 3a,b} described by:
\begin{equation}
    \text{PL}(t) = A_1 e^{-\frac{t}{\tau_1}} + A_2 e^{-\frac{t}{\tau_2}} + A_3 e^{-\frac{t}{\tau_3}}
\end{equation}
Where $A_i$ and $\tau_i$ are the respective amplitudes and lifetimes which are shown for all samples in \textbf{Table \ref{stable:perov-trpl}} and \textbf{\ref{stable:trpl-bilayer}} alongside the average lifetimes $\bar{\tau}$ calculated as:
\begin{equation}
    \bar{\tau} = \sum\limits_{i=1}^3 A_i \tau_i
\end{equation}

\begin{table*}[!ht]
  \centering
  \caption{Photoluminescence lifetimes of perovskite emission ($\lambda > 725$\,nm) extracted via triexponential fitting for all neat perovskite films.}
  \begin{tabular}{|>{\columncolor[gray]{0.9}}c|>{\columncolor[gray]{0.95}}c|>{\columncolor[gray]{0.9}}c|>{\columncolor[gray]{0.95}}c|>{\columncolor[gray]{0.9}}c|>{\columncolor[gray]{0.95}}c|>{\columncolor[gray]{0.9}}c|>{\columncolor[gray]{0.95}}c|}
    \hline
    \textbf{Sample} & $A_1$ & $\tau_1$ (ns) & $A_2$ & $\tau_2$ (ns) & $A_3$ & $\tau_3$ (ns) & $\bar{\tau}$ (ns)\\
    \hline
    Control & 0.60 & 8.9 & 0.35 & 45 & 0.05 & 199 & 31.0 \\
    P1 & 0.47 & 12.4 & 0.41 & 72 & 0.12 & 219 & 61.6 \\
    P4 & 0.37 & 17 & 0.44 & 77 & 0.19 & 271 & 91.7 \\ 
    P8 & 0.80 & 8.1 & 0.19 & 28 & 0.01 & 141 & 13.2 \\
    \hline
  \end{tabular}
  
  \label{stable:perov-trpl}
\end{table*}

\begin{table*}[!ht]
  \centering
  \caption{Photoluminescence lifetimes of perovskite emission ($\lambda > 725$\,nm) extracted via triexponential fitting for all upconversion samples.}
  \begin{tabular}{|>{\columncolor[gray]{0.9}}c|>{\columncolor[gray]{0.95}}c|>{\columncolor[gray]{0.9}}c|>{\columncolor[gray]{0.95}}c|>{\columncolor[gray]{0.9}}cl|>{\columncolor[gray]{0.95}}c|>{\columncolor[gray]{0.9}}c|>{\columncolor[gray]{0.95}}c|}
    \hline
    \textbf{Sample} & $A_1$ & $\tau_1$ (ns) & $A_2$ & $\tau_2$ (ns) & $A_3$ & $\tau_3$ (ns) & $\bar{\tau}$ (ns)\\
    \hline
    Control/Rub & 0.52 & 10.3 & 0.45 & 49 & 0.03 & 148 & 31.8 \\
    P1/Rub & 0.60 & 13.7 & 0.35 & 74 & 0.05 & 260 & 47.1 \\
    P4/Rub & 0.52 & 13.2 & 0.41 & 60 & 0.07 & 204 & 45.7 \\
    P8/Rub & 0.71 & 12.2 & 0.26 & 43 & 0.03 & 189 & 25.5 \\
    \hline
  \end{tabular}
  \label{stable:trpl-bilayer}
\end{table*}

\clearpage

\section{Supporting Note 4}

The non-Lorentzian functions fit to the data points in \textbf{Figure 3c} is described by:
\begin{equation}
    f(B) = A \cdot \frac{B^2}{(|B|+\sigma)^2}
\end{equation}
Where $A$ describes the amplitude and $\sigma$ describes the sharpness of the curve.\\

\begin{figure}[!ht]
    \centering
    \includegraphics[width=\linewidth]{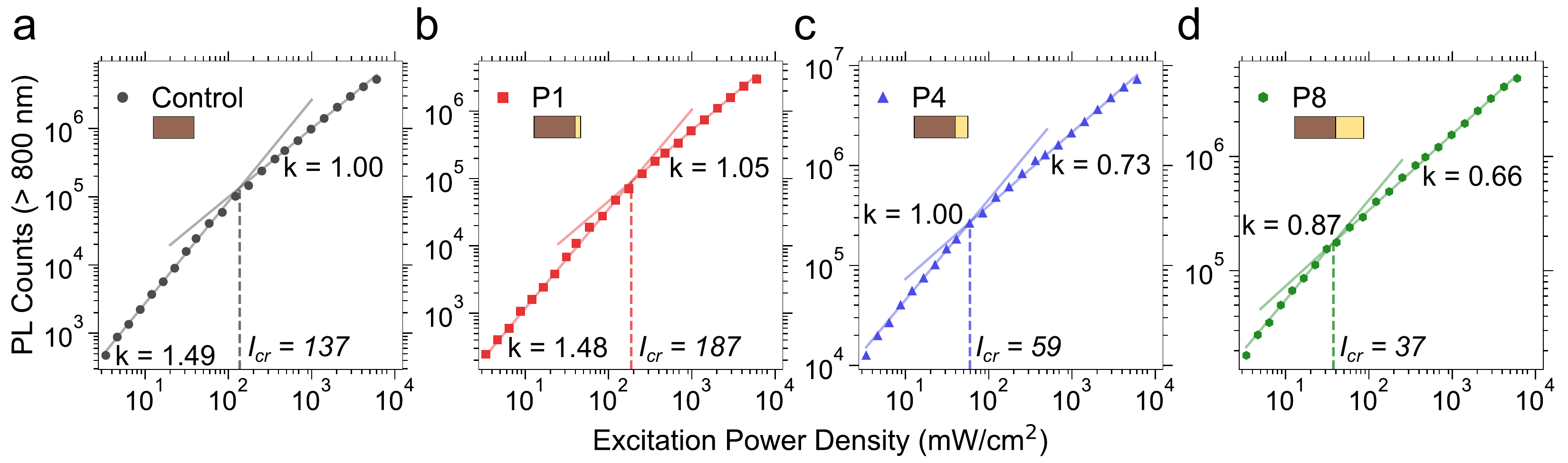}
    \caption{Excitation dependency of the emission ($\lambda > 800$\,nm for the neat perovskite films a) Control, alongside the PEAI-treated samples b) P1, c) P4, and d) P8. Linear fits and the extracted gradients are overlaid, additionally where there is a regime change the crossover intensity ($I_{cr}$) value is denoted.}
    \label{sfig:pdepperov}
\end{figure}

\section{Supporting Note 5}

The power dependence of the perovskite emission ($> 800$\,nm) is shown in \textbf{Supplementary \ref{sfig:pdepperov}}. Both the Control/Rub bilayer and P1/Rub trilayer show typical excitation dependencies with exhibiting crossover thresholds ($I_{cr}$) and gradients approximately taking on values of multiples of $1/2$.\textsuperscript{4,5} However, the two trilayers with the thicker 2D perovskite capping layers show reduced gradients as well as reduced crossover regions potentially due to the complex energetic environment within the 2D perovskite spacer layer as seen in \textbf{Figure \ref{sfig:perov-pl}}.\\

\begin{figure}[!ht]
    \centering
    \includegraphics[width=\linewidth]{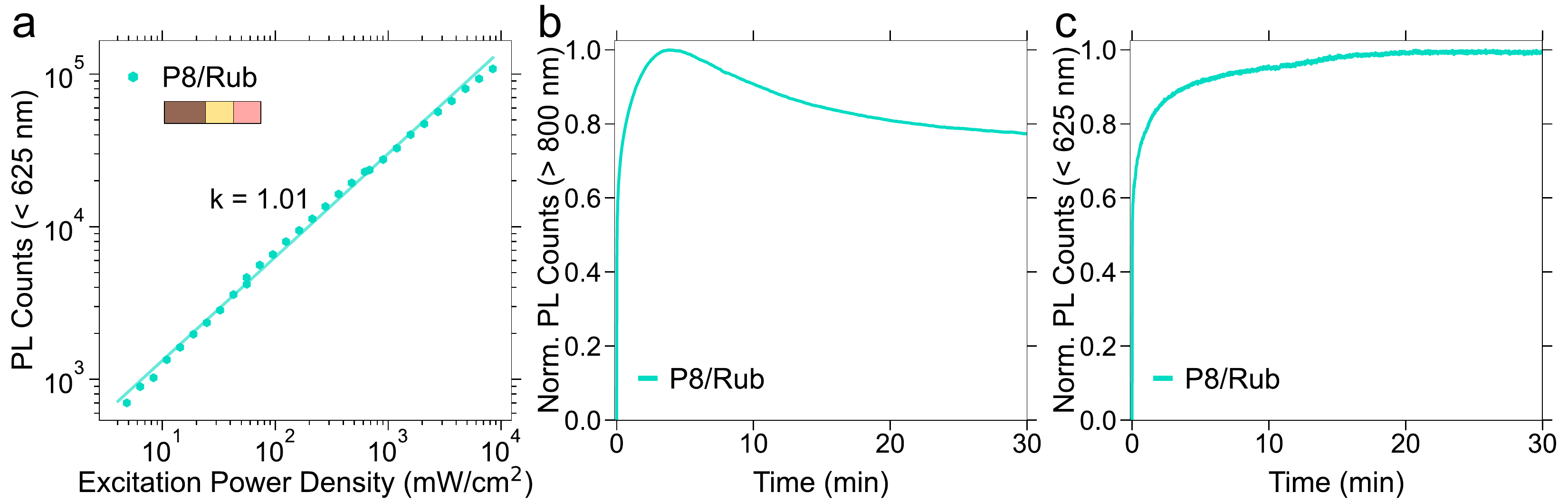}
    \caption{(a) Excitation dependence of the upconverted emission ($\lambda < 625$\,nm) for the P8/Rub trilayer. A Linear fit and the extracted gradient is overlaid. (b) Normalised photoluminescence of the bulk perovskite for the P8/Rub trilayer over 30 minutes. (c) Normalised upconverted photoluminescence over 30 minutes for the P8/trilayer.}
    \label{sfig:p8pdep}
\end{figure}

\clearpage

\section*{References}

\begin{enumerate}
  \renewcommand{\labelenumi}{(\theenumi)}
  \item Jacobsson, T. J.; Correa-Baena, J. P.; Halvani Anaraki, E.; Philippe, B.; Stranks, S. D.; Bouduban, M. E.; Tress, W.; Schenk, K.; Teuscher, J.; Moser, J. E.; Rensmo, H.; Hagfeldt, A. Unreacted PbI2 as a Double-Edged Sword for Enhancing the Performance of Perovskite Solar Cells. \textit{Journal of the American Chemical Society} \textbf{2016}, 138, 10331–10343.
  \item Jiang, Q.; Zhao, Y.; Zhang, X.; Yang, X.; Chen, Y.; Chu, Z.; Ye, Q.; Li, X.; Yin, Z.; You, J. Surface passivation of perovskite film for efficient solar cells. \textit{Nature Photonics} \textbf{2019}, 13, 460–466.
  \item Fu, Y.; Zheng, W.; Wang, X.; Hautzinger, M. P.; Pan, D.; Dang, L.; Wright, J. C.; Pan, A.; Jin, S. Multicolor Heterostructures of Two-Dimensional Layered Halide Perovskites that Show Interlayer Energy Transfer. \textit{Journal of the American Chemical Society} \textbf{2018}, 140, 15675–15683.
  \item Schmidt, T.; Lischka, K.; Zulehner, W. Excitation-power dependence of the near-bandedge photoluminescence of semiconductors. \textit{Physical Review B} \textbf{1992}, 45, 8989–8994.
  \item Spindler, C.; Galvani, T.; Wirtz, L.; Rey, G.; Siebentritt, S. Excitation-intensity dependence of shallow and deep-level photoluminescence transitions in semiconductors. \textit{Journal of Applied Physics} \textbf{2019}, 126, 175703.
\end{enumerate}

\end{document}